\documentclass[%
 aip,
 amsmath,amssymb,
reprint,%
]{revtex4-1}

\usepackage{graphicx}
\usepackage{dcolumn}
\usepackage{bm}
\usepackage[utf8]{inputenc}
\usepackage[T1]{fontenc}
\usepackage{mathptmx}
\usepackage{mathrsfs}
\usepackage{etoolbox}
\usepackage{lipsum}
\usepackage{color,soul}
\usepackage[normalem]{ulem}
\usepackage{xspace}
\usepackage{xcolor}
\usepackage{hyperref}
\usepackage{braket}
\usepackage{placeins}

\newcommand{\inse}{In$_2$Se$_3$}

\newcommand{\qsgw}{QS$GW$\xspace}
\newcommand{\qsgwh}{QS$G\widehat{W}$\xspace}

\makeatletter
\def\@email#1#2{%
 \endgroup
 \patchcmd{\titleblock@produce}
  {\frontmatter@RRAPformat}
  {\frontmatter@RRAPformat{\produce@RRAP{*#1\href{mailto:#2}{#2}}}\frontmatter@RRAPformat}
  {}{}
}%
\makeatother

\begin{document}

\preprint{ACO25-AR-00343}

\title{Many-body description of two-dimensional van der Waals ferroelectric $\alpha-$\inse}

\author{Denzel Ayala}
\altaffiliation[Also at ]{National Laboratory of the Rockies, Golden, Colorado 80401, USA}
\affiliation{Department of Physics, University at Buffalo, State University of New York, Buffalo, New York 14260, USA
}
\email{denzelay@buffalo.edu}

\author{Dimitar Pashov}
\affiliation{Department of Physics, King's College London, London WC2R 2LS, United Kingdom}

\author{Tong Zhou}
\affiliation{Eastern Institute for Advanced Study, Eastern Institute of Technology, Ningbo, Zhejiang 315200, China}

\author{Kirill Belashchenko}
\affiliation{Department of Physics and Astronomy, University of Nebraska-Lincoln, Lincoln, Nebraska, 68588, USA}

\author{Mark van Schilfgaarde}
\affiliation{National Laboratory of the Rockies, Golden, Colorado 80401, USA}
\email{Mark.vanSchilfgaarde@nrel.gov}

\author{Igor \v{Z}uti\'{c}}
\affiliation{Department of Physics, University at Buffalo, State University of New York, Buffalo, New York 14260, USA}
\email{zigor@buffalo.edu}

\date{\today}
\begin{abstract}
Two-dimensional (2D) van der Waals ferroelectrics are recognized for enabling many applications, from memory and logic to neuromorphic computing, as well as transforming other materials to control electronic phase transitions and topological states.  While these materials are typically weakly correlated and expected to have their ground-state properties well described with the commonly used density functional theory, by focusing on bilayers and trilayers of In$_2$Se$_3$ we show that this approach may not be reliable. The underlying electronic structure strongly depends on the polarization structure of the multilayer system and is surprisingly challenging to accurately calculate, requiring a high-fidelity many-body theory of the quasiparticle self-consistent \textit{GW} approximation.  We develop this underlying description by extending the capabilities of Green function implementation within the open-source Questaal package. We show that even a sophisticated hybrid functional approach may fail to predict a nonvanishing gap in a bilayer In$_2$Se$_3$ and yields charge density, polarization, and band offsets that strongly deviate from the many-body picture. We discuss the implications of these computational advances for future opportunities in 2D ferroelectrics.
\begin{description}
\item[KEYWORDS]
Two-dimensional van der Waals ferroelectrics, many-body perturbation theory
\end{description}
\end{abstract}

\maketitle

\section{Introduction}

\begin{figure*}
    \centering
    \includegraphics{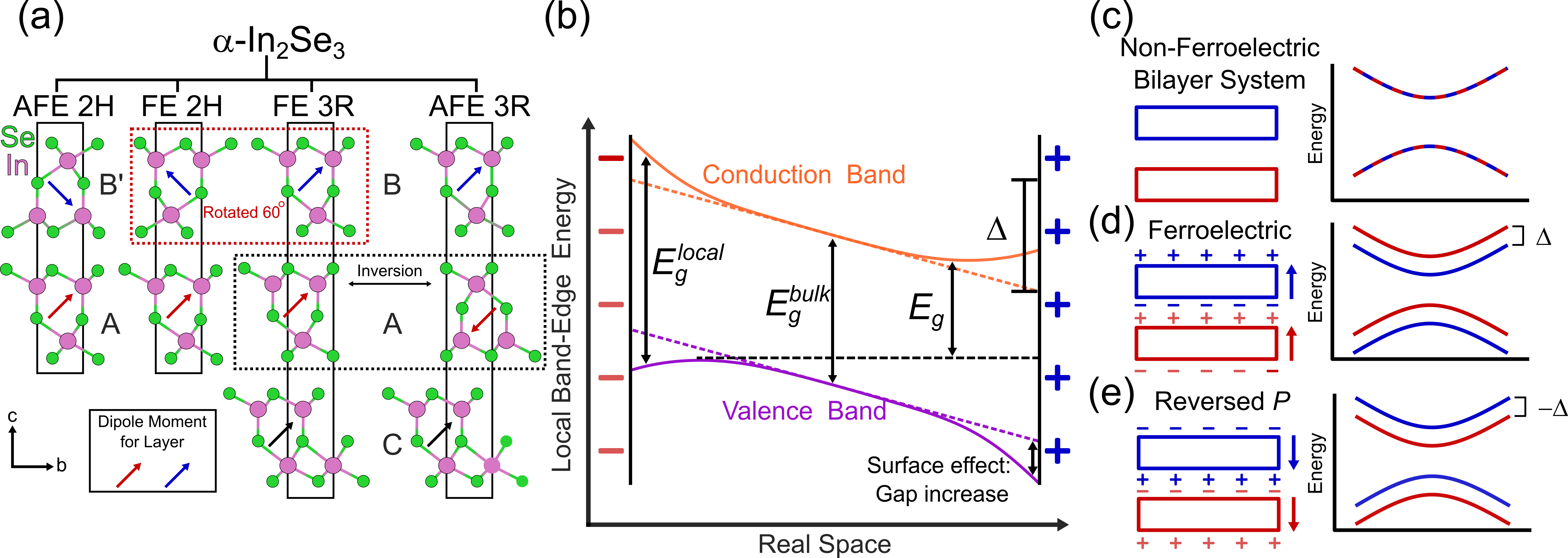}
    \caption{(a) Stacking configurations of $\alpha$-\inse: AB'A hexagonal 
    (2H) and ABC rhombohedral (3R) in both out-of-plane (OOP) ferroelectric (FE) and OOP antiferroelectric (AFE) systems.  
    In the 3R configuration the three layers are related by an in-plane translation relative to layer A. However, in the 2H configuration, layer B' is also rotated by 60$^\circ$ about the c-axis. As the 2H is a two layer structure, the in-plane dipole moment alternates between layers. (b)  A two-channel model of the spatially-resolved band structure inside a capacitor in  the electric field generated by the ferroelectricity. The electronic potential is decomposed into a periodic part and the uniform field from the capacitor. The band edges are depicted as varying linearly with position in the FE region. Each surface forms a local band gap, $E_g^\mathit{local}$,  influenced by multiple factors that can be lumped together as surface effects. For a sufficiently thick sample, there is a bulk gap, $E_g^\mathit{bulk}$, that can be computed using a periodic system. For thin or finite systems the global gap, $E_g$, is the energy difference between the highest valance band and the lowest conduction band. (c){-}(e) Schematic projections into the top and bottom surfaces of the frontier bands for different FE states. (c) Non-FE vdW bilayer with identical top and bottom surfaces and (d) 
    with a ferroelectrically induced electrostatic shift, $\Delta$, between the top and bottom surface.  (e) The sign of $\Delta$ is inverted with the reversal of OOP  polarization, $P$. 
    }
    \label{fig:CF1}
\end{figure*}

    Ferroelectric (FE) materials have a broken inversion symmetry and two or more (meta)stable polarization states that can be switched by applying an electric field.\cite{Waser:2012} These properties are both valuable as inherent to FEs and to how they could transform other materials.\cite{Scott2015:S,Fei2025:S,Lee2015:S,Tao2021:JPDAP,Varotto2021:NE,Yeh1968:PSS,Ibrahim2109:2DM} FEs enable greater carrier density changes and  electric fields, ${\bm E}_\text{ext}$, than what is possible  even by ion-liquid gating with $E_\text{ext}\lesssim0.5\;$V/\AA, approaching typical values to break the chemical bond, four orders of magnitude larger than the breakdown of air.\cite{Lazic2016:PRB,Ahn2006:RMP} Therefore, FEs can extend prior work on electronic phase transitions, inducing ferromagnetism and superconductivity, or controlling topologically-nontrivial states.\cite{Ahn2006:RMP,Denisov2025:PRL,Chen2024:AFM,Liang2023:NE}  The control of (anti)ferroelectricity also provides a powerful method to realize versatile altermagnets with vanishing magnetization and tunable spin polarization.\cite{Duan2025:PRL, Gu2025:PRL,Zhu2025:NL,Camerano2025:npj2DMA,Zhu2025:SCPMA,Mavani2025:PRB,Zhu2025:Xb}

    Since the discovery of molecular FEs,\cite{Valasek1921:PR} ferroelectricity and its nonvolatile polarization, has been found in many classes of materials and sought for various applications.\cite{Mikolajick2021:JAP,Rao2022:PRM,Wang2022:C} While for conventional bulk FEs a simple phenomenological description is often sufficient,\cite{Waser:2012} there is a growing interest in two-dimensional (2D) van der Waals (vdW) materials and heterostructures which  pose a challenge to accurately describe the electronic structure of vdW FEs from first principles.\cite{Wang2017:2DM,Guan2019:AEM} These FEs may overcome the high-leakage current,\cite{Sun2010:APL,Feng2025:IEEETED,Yan2008:JAP,Wang2025:NC} surface dangling bonds,\cite{Qi2021:AM,Ho2013:ACSAMI,Dutta2021:NPJ2DM} and complex fabrication\cite{Wang2025:NC,Gundlapudi:2023,Qiu2023:SR,Zhang2015:PMS,Li2025:MM} that plague  their conventional counterparts. 

    The need to accurately capture electronic structure of vdW FEs is crucial beyond the vdW monolayers where, for example, failing to obtain an accurate band gap could be subsequently corrected by using the experimental values. With our focus on bilayer and trilayer vdW FEs, other considerations, such as band alignment and hybridization, become relevant; although often difficult to access experimentally, they are essential to understand how materials properties may be tuned by proximity effects,\cite{Zutic2019:MT,Zhang2021:NL,Huang2022:NL,Zhu2025:Xa} by changing the number of layers and their relative twist angles,\cite{Wu2015:NL} or by introducing strain.\cite{Felton2025:NC,Zhang2022:JPCL,Hu2017:RSCA} The first step to long-range interactions that may influence electronic structure is describing weak and dispersive vdW interlayer forces. For example, commonly used local exchange functionals predict no bonding for bilayer graphene.\cite{Lazic2016:PRB} 

    In this work, we  study the electronic structure of vdW FEs from first principles on the example of In$_2$Se$_3$, illustrated in Fig.~\ref{fig:CF1}(a). Since it is a weakly correlated $s-p$ material, it is expected that common density functional theory (DFT)\cite{Payne1992:RMP} is an appropriate choice with its low computational cost and reasonable predictive power, which was used both for In$_2$Se$_3$ and its vdW heterostructures.\cite{Han2023:RSC,Han2023:SR,Eom2023:NC,Liao2021:APL,Zhang2024:NPJCM} Remarkably, by extending a versatile open-source package Questaal\cite{Pashov2020:CPC} using the implementation of Green functions through the many-body perturbation theory, we reveal that a reliable determination of even seemingly simple ground state properties of In$_2$Se$_3$ becomes very difficult. This is surprising as it is the case where the DFT approach is expected to work well. 

    The implications of our findings are then twofold: (i) To the best of our knowledge, this is the first many-body study of FEs  within Questaal as for their accurate description we needed to include the dipole correction\cite{Neugebauer1992:PRB} from first principles. This now opens many opportunities for the subsequent studies by taking advantage of the Questaal's powerful framework building on the improved self-consistent quasiparticle \textit{GW} approximation and Bethe-Salpeter equation
    .\cite{Pashov2020:CPC,Cunningham2023:PRB} (ii) There is a growing interest in vdW FEs and, in particular, In$_2$Se$_3$, whose out-of-plane (OOP) polarization has found use in neuromorphic computing,\cite{Shin2025:AEM,Baohua2025:JAC,Mohta2021:RSCA} optoelectronics,\cite{Wang2023:AS,Jia2023:ACSN,Wang2024:AIPA} nanoelectronics,\cite{Liu2025:NR,Kim2025:S,Wan2019:AFM} and spintronics,\cite{Tao2017:PRB,Jafari2022:PRM,Milivojevic2024:2DM,Milivojevic2025:MF} as well as offering important advantages in transforming other materials.\cite{Zhang2021:NL,Huang2022:NL} However, it is critical to elucidate the limitations of the DFT approach for vdW FEs and whether they can be overcome by its more accurate implementations, such as by using hybrid functionals.\cite{Krukau2006:JCP}

    In Section II we provide some background discussion of the computational methods employed. This includes both common DFT approaches and different levels of the \textit{GW} approximation, where the enhanced accuracy of the self-consistent quasiparticle \textit{GW} implementation can be crucial in describing vdW FEs. In Section III we discuss the main results of this work and show how the electronic structure depends on the choice of the first-principles methods. Here we also consider a qualitative capacitor model from Fig.~\ref{fig:CF1}(b) and observe that the electronic structure depends strongly on the polarization configuration in the multilayer structure, as depicted in Figs.~\ref{fig:CF1}(c)-\ref{fig:CF1}(e). In Section IV we discuss further challenges as well as opportunities made possible by these advances in the accurate description of vdW FEs.
    
\section{Computational Methods}

\subsection{Theoretical Framework} 

    Electronic structure methods usually approximate the bare Coulomb interaction, $v$, by decomposing into classical and quantum components. The local classical components include the external potential, $V_\text{ext}({\bf r})$, that acts as a fixed background, and the electrostatic contribution from electrons as the Hartree potential, $V_H({\bf r})$, where ${\bf r}$ is the relative coordinate. While these two contributions stay consistent across the different approximations, the level of the employed theory is defined by the specific treatment of the quantum mechanical exchange and correlation components, $V_\text{xc}(\mathbf{r},\mathbf{r'})$. This can be summarized with a Schr\"odinger-like equation 
    \begin{gather}
        \left[ -\frac{\hbar^2}{2m}\nabla^2 + V_\text{ext}(\mathbf{r}) + V_H(\mathbf{r}) + V_\text{xc}(\mathbf{r},\mathbf{r'}) \right] \Psi_i = \varepsilon_i \Psi_i, 
        \label{eq:schrodinger}
    \end{gather}
    where $\Psi_i \equiv \Psi_i(\mathbf{r})$ is the wave function of the electronic state $i\equiv(\mathbf{k},n)$ with $\mathbf{k}$ and $n$ the wave vector and band index, respectively, while $\varepsilon_i$ is the corresponding eigenvalue. It is convenient to employ atomic Rydberg units, $\hbar=2m=e^2/2=1$, where $\hbar$  is the Planck's constant, while $m$ and $e$ are the electron's mass and charge.
    
    In DFT, within the Kohn-Sham construction,\cite{Kohn1965:PR} the one-body density, $n(\mathbf{r})$, is obtained from the solutions of a fictitious one-body Hamiltonian $H_0$, whose effective local potential, $V_\text{eff}(\mathbf{r})$, is the functional derivative of the potential term in the total energy\cite{Payne1992:RMP}
    \begin{align}
        V_\text{eff}(\mathbf{r}) & =  V_\text{ext}(\mathbf{r}) + \frac{\delta E_H[n(\mathbf{r})]}{\delta n(\mathbf{r})}  + \frac{\delta E_\text{xc}[n(\mathbf{r})]}{\delta n(\mathbf{r})} \nonumber \\
        & \equiv V_\text{ext}(\mathbf{r}) + V_H([n(\mathbf{r})],\mathbf{r}) + V_\text{xc}([n(\mathbf{r})],\mathbf{r}). 
        \label{eq:dft}
    \end{align}
    However, Eq.~(\ref{eq:dft}) is subject to two types of errors. (i) The functional describing correlations from electron-electron interactions is unknown, and an uncontrolled ansatz is used instead. (ii) The eigenvalues of this fictitious one-body Schr{\"o}dinger equation have no physical meaning, even though they are often interpreted as electronic excitations.  The functional is only designed to describe the ground state. Even if it was exact, the eigenvalues need not correspond to the excitation energies. They resemble correlation energies, but their fidelity is much less than  for the  ground-state properties.\cite{Gruning2006:JCP} 
    
    To retain the simplicity and efficiency of DFT, a patchwork of improvements have been put forward. Usually, they add some nonlocality, $V_\text{eff}(\mathbf{r})\rightarrow V_\text{eff}(\mathbf{r},\mathbf{r'})$, making the potential orbital dependent. Some popular choices are to use a hybrid admixture of the DFT energy functional and the Fock exchange $E_x^{HF}$:\cite{Adamo1999:JCP} 
    \begin{gather}
        E_\text{xc}^\text{Hybrid} = \alpha E_x^{HF}(\mathbf{r},\mathbf{r'}) + (1-\alpha)E_x^{DFT} + E_c^{DFT}, 
        \label{eq:hybrid}
    \end{gather}
    possibly admixing a range-truncated Hartree-Fock (HF) approximation,\cite{Krukau2006:JCP} adding a Hubbard \textit{U} to the functional,\cite{Hubbard1963:PRSL} or including new terms intended to capture the $r^{-6}$ spatial decay of the vdW interaction.  While all of these extensions can improve some particular properties, they all are \textit{ad hoc} with adjustable parameters, like $\alpha$ in Eq.~(\ref{eq:hybrid}), which makes the underlying computational framework difficult to improve in a systematic manner. Each of these parameterized Hamiltonians attempt to induce nonlocality in the potential. However, the physical basis of each type of approximation is different and, absent a high-level theory, it is not possible to justify why one DFT extension should be preferred over another. 
    
    Many-body perturbation theory is a diagrammatic method that can overcome the uncontrolled approximations in DFT. Hedin\cite{Hedin1965:PR} constructed a formally exact diagrammatic expansion for the one-particle Green function $G_0(\mathbf{r},\mathbf{r'})$, which is a useful concept for describing correlation effects, and whose screened potential is an explicit functional of the Green function.  The diagrammatic expansion is formulated in powers of the screened Coulomb interaction \textit{W}, where $W = \epsilon^{-1} v$, and $\epsilon^{-1}(\mathbf{r},\mathbf{r'},\omega)$ is an inverse dielectric function, and $\omega$ is an angular frequency. The lowest-order diagram (often the only one employed) is the ``\textit{GW}'' approximation while ``\textit{Gv}'' is equivalent to the unscreened Hartree-Fock theory. This approximation is already rather good because screening is by far the most important many-body effect. To connect the many-body problem to an effective single-particle problem\cite{Bechstedt:2016} is useful to introduce the self-energy as an effective exchange-correlation potential
    \begin{gather}
        \Sigma(\mathbf{r},\mathbf{r'},\omega) =\frac{i}{2\pi}\int d\omega' G_0(\mathbf{r},\mathbf{r'},\omega-\omega') W(\mathbf{r},\mathbf{r'},\omega')e^{-i\delta \omega}, 
        \label{eq:selfenergy}
    \end{gather}
    which is nonlocal in both space and time.

    However, the \textit{GW} approximation has different  shortcomings. It is a perturbation theory so the result depends on its starting one-body Hamiltonian's effective potential
    \begin{gather}
        \Delta V(\omega) =  V_\text{ext}(\mathbf{r}) + V_H(\mathbf{r}) + \Sigma(\mathbf{r},\mathbf{r'},\omega) - V_\text{eff}(\mathbf{r}),
        \label{eq:perturbation}
    \end{gather}
    DFT is by far the most common starting point, but it is not unique. Also, even within DFT there are as many results as there are flavors of DFT.  Thus, single-shot \textit{GW} based on DFT ($G^\mathrm{DFT}W^\mathrm{DFT}$) contains the same ambiguities as its starting point does. Furthermore, \textit{GW} is not a conserving approximation. For example, there is an ambiguity in the charge density for $N$ electrons as the electron density $n(\mathbf{r})$ used to generate \textit{GW} through $H_{0}$ and the one-body density matrix    
    \begin{gather}
        \rho(\mathbf{r},\mathbf{r'})  = N\sum_i w_i \int d\mathbf{r}_2 . . .d\mathbf{r}_N \Psi^*_i(\mathbf{r},...,\mathbf{r}_N) \Psi_i(\mathbf{r'},...,\mathbf{r}_N), 
        \label{eq:density}
    \end{gather}
    generated by \textit{GW} are not equivalent by construction. Here  $w_i$ is the probability of state $i$ in the density distribution. Usually, the ambiguity is ignored because only the diagonal elements of the density matrix are calculated, i.e., $\rho(\mathbf{r},\mathbf{r'})\delta(\mathbf{r}-\mathbf{r'}) \equiv n(\mathbf{r})$;\cite{Martin:2016} keeping the density fixed. However, the off-diagonal elements are crucial in this system. They can capture quantum-mechanical effects and correlations, including off-diagonal long-range order, known before DFT was introduced.\cite{Penrose1956:PR,Bechstedt:2016,Hybertsen1986:PRB} 
    
    Both non-uniqueness and the non-conserving nature can be circumvented by self-consistency; however, fully self-consistent \textit{GW} has long been known to be a poor approximation in solids, even for the electron gas.\cite{Holm1998:PRB}  The main source of error can be traced to the lack of cancellation of the quasiparticle renormalization factor $Z$; see Appendix A of earlier work.\cite{Kotani2007:PRB} To circumvent this, a ``quasiparticlized'' form of the self-consistency was developed, QS\textit{GW},\cite{Faleev2004:PRL,Schilfgaarde2006:PRL} which can be represented by the following equation for the ``dressed quasiparticles''  
    \begin{gather}
        \left[  -\frac{1}{2}\nabla^2 + V_{ext}(\mathbf{r}) + V_H(\mathbf{r}) + \text{Re}[\Sigma(E_i)] - E_i \right] \Phi_i = 0, 
        \label{eq:quasiparticleCondition}
    \end{gather}
    where $\Phi_i(\mathbf{r})$ is the QP eigenfunction and $E_i$ its energy. 

    In this self-consistency procedure, the one-body Hamiltonian, $H_0^\text{DFT}$, is replaced by the QP counterpart, $H_0^\text{QP}$, which is repeatedly updated. This update is partially accomplished by replacing the DFT $V_\text{xc}$ by the Hermitian part of the self-energy which is taken to be static and evaluated around the Fermi level, $\Sigma_0$. By implementing this self-consistency, QS\textit{GW} no longer relies on DFT. QS\textit{GW} removes the starting point dependence and, by minimization of a variational property,\cite{Ismail2017:JPCM} the quasiparticlized self-energy, has the same poles as the Hermitian part of the dynamical one. Thus, the obtained $H_0^\text{QP}$ and its corresponding QP levels can be interpreted as excitation energies and eigenfunctions that have physical meaning as well. Furthermore, QS\textit{GW} uncovers systematic errors inherent in the \textit{GW} approximation, which are not evident in $G^\mathrm{DFT}W^\mathrm{DFT}$. In particular, the screening is insufficiently described by the time-dependent Hartree approximation of $\epsilon^{-1}$.\cite{Hedin1965:PR} A recent work showed that improving $\epsilon^{-1}$ by the addition of ladder diagrams greatly reduces the systematic error in QS\textit{GW}.\cite{Cunningham2023:PRB} We use the related notation $G\widehat{W}$\cite{Cunningham2023:PRB} to indicate calculations which include the  contributions of ladder diagrams in \textit{W}. Remarkably, QS$G\widehat{W}$ defines, with high fidelity, a description of excitations in many different types of insulators, without {\em any} adjustable parameters.\cite{Pashov2020:CPC}

    The electrostatic field is built through screening the $V_\text{ext}$, which includes the nuclear potential from the electron's perspective.  Screening is described through the polarizability, as a part of the dielectric function $\epsilon$.  How the screening is built up differs in many-body perturbation theory (MBPT) compared to mean-field theories such as HF, DFT, and their combination in hybrid functionals. The irreducible polarizability $P$ is not calculated directly in these mean-field theories but appears implicitly in the self-consistency cycle: a trial input density $n^0$ generates an input potential $v^0$ and output density $n^1$, which, in the next iteration, generates $n^2$, and so on until self-consistency is reached. In practice, input-output pairs are mixed to control convergence.  At each iteration, the change in the density $\delta n^{i+1}=n^{i+1}-n^{i}$ generates the change in the input potential $\delta v^{i+1}$, which may be thought of as a perturbation. If the \textit{exact} dielectric function were known, the exact $\delta n$ (to linear order) could be calculated from $\epsilon^{-1}\delta n$.  Since $\delta v$ would also be exact, implementation of self-consistency would not be needed (to linear order). Instead, screening is built up through self-consistency in a mean-field theory, with each $\delta n^{i+1}$ updating $n$ towards a converged screened form. In contrast, the screening is calculated explicitly in MBPT, with $P$ written as\cite{Martin:2016,Aryasetiawan1998:RPP}
    \begin{equation}
        P(\mathbf{r},\mathbf{r}',\omega)\equiv -\,i\,\frac{\delta n(\mathbf{r},\omega)}{\delta v_{\mathrm{tot}}(\mathbf{r}',\omega)},
        \label{eq:defP}
    \end{equation}
    where $v_\text{tot}$ is the total potential acting on the electrons. In MBPT, $\Sigma$ is generated explicitly from $P$; in a mean-field theory, $P$ appears as a byproduct of an assumed, material-agnostic form for the exchange-correlation potential.  Additionally, $P$ and $\Sigma$ are dynamical in MBPT [see Eq.~(\ref{eq:selfenergy})], while both are assumed to be static in mean-field theories [see Eq.~(\ref{eq:dft})].

    An important difference between QS\textit{GW} and \textit{GW} based on one-shot perturbation from DFT or a hybrid functionals may be understood as a feedback between $P$ and $\Sigma$: changes in $\Sigma$ modify $P$ (notably $P$ increases rapidly as the band gap closes), which modifies $\Sigma$, etc.  This feedback is only sensed through self-consistency, where $P$ is updated with $\Sigma$.  Screening in a mean-field theory is not material-specific but tied to an ansatz, so $GW$ as a perturbation to it does not properly account for it.  These observations were highlighted in another system with similar properties and elements, $\mathrm{CuIn}(\mathrm{S},\mathrm{Se})_{2}$,\cite{Vidal2010:PRL} as well as in TiSe$_{2}$.\cite{Acharya2021:NPJCM}

    Finally, how well $P$ itself is calculated depends on the level of theory in MBPT.  In the random phase approximation (RPA), on which \textit{GW} is usually based, $P$ is calculated in the independent-particle approximation, as the tensor product $iG{\otimes}G$.\cite{Martin:2016,Onida2002:RMP}  This misses excitonic effects, making $P$ too small. $P$ can be greatly improved by adding ladder diagrams~\cite{Cunningham2023:PRB}; they can be especially important in wide-gap systems, but we have found they are rather small for systems we study here.

    To see intuitively how the previously noted feedback appears, consider the RPA polarizability $iG{\otimes}G$ written~\cite{Onida2002:RMP} in terms of eigenfunctions $\Psi_i$ and eigenvalues $\varepsilon_i$
    \begin{equation}
        P_\text{IP}(\mathbf{r},\mathbf{r}',\omega)=\sum_{{i=occ}\atop{j=unocc}}
        \frac{\Psi_i(\mathbf{r})\Psi_j^*(\mathbf{r})\,\Psi_j(\mathbf{r}')\Psi_i^*(\mathbf{r}')}{\omega-(\varepsilon_j-\varepsilon_i)+i\eta},
        \label{eq:RPA_Polarizability}
    \end{equation}
    where $\eta\rightarrow 0^+$ enforces causality. $P(\omega{\rightarrow}0)$ is sensitive to  the smallest value of $\Delta\varepsilon_{ij}=\varepsilon_j-\varepsilon_i$ (the gap) and diverges when it vanishes.  If $P$ is too large (small) the gap it generates will be too small (large).  If $P$ (or its proxy) is poorly calculated, the gap will be as well. Similarly, errors in the gap generate errors in $P$.  When ladder diagrams are added to $P$, it becomes more accurate for the bands generating it; self-consistency ensures that the energy bands are consistent with $P$.

    This background discussion of several different methods and approximations will allow us to better understand an unexpected need for high-fidelity first-principles description of ferroelectricity in 2D vdW materials.

\subsection{Computational Implementation} 
    
    In our computational studies we examine different FE configurations of  a monolayer (1L), bilayer (2L), trilayer (3L), and bulk $\alpha$-In$_{2}$Se$_{3}$, as well as its 2H and 3R polytypes [Fig.~\ref{fig:CF1}(a)], showing how electronic structure varies with different dipole orderings for different approximations to $H_0$. In particular, we compare two common density functionals: Local density approximation (LDA)\cite{Kohn1965:PR,vonBarth1972:JPCSSP} and the generalized gradient approximation (GGA) as implemented by Purdue-Burke-Ernzerhof (PBE),\cite{Perdew1996:PRL,Adamo1999:JCP} as well as using a generally more accurate hybrid functional Heyd–Scuseria–Ernzerhof (HSE06).\cite{Krukau2006:JCP} These implementations then provide a valuable reference for our results obtained from \qsgw, \qsgwh, and single-shot $GW$, evaluated under the LDA density,  ($G^\mathrm{LDA}W^\mathrm{LDA}$).

    All the FE zinc-blende $\alpha$-\inse\ structures were taken from Rietveld refinements of powder X-ray diffraction data\cite{Kupers2018:IC} and used without relaxation. At the GGA level the differences between the relaxed and unrelaxed 3R 2L were minimal. While the OOP dipole moment per unit area quantitatively decreased by $\sim5\%$ when the structure was relaxed, the  dispersion of the bands and the band alignment were unchanged. For thin film system (e.g. bilayers and trilayers), a vacuum layer at least 80$\;$\AA\ thick was used to avoid interactions between periodic slabs.  We implemented a dipole correction\cite{Neugebauer1992:PRB} at the DFT level for the first time implemented in Questaal and applied it for selected 2L calculations to eliminate the electric field in the vacuum.
    
\subsubsection{VASP Calculations}
    In the Vienna Ab-initio Simulation Package (VASP),\cite{Kresse1999:PRB,Kresse1996:PRB, Kresse1996:CMS} we use GGA and HSE06 fuctionals. Both simulations were performed using the projector-augmented wave (PAW) pseudopotential with a plane wave cutoff energy of $500\;$eV. The Brillouin-zone integration for the  GGA (HSE06) calculation was performed using $13\times13\times1$ $(13\times13\times1)$ Gamma-centered mesh with a Gaussian smearing of $0.1\;$eV. The total energy is converged to less than $1\times10^{-6}\;$eV. For GGA we tested the rev-vdW-DF2 functional\cite{Hamada2014:PRB} for the determination of the effective potential and electronic properties, but it had negligible effects and was omitted in the reported results.

\subsubsection{QS$GW$ Calculations}
   
    Our all-electron, augmented-wave + muffin-tin orbital implementation of \qsgw was described in detail in Kotani et. al.'s work.\cite{Kotani2007:PRB} For the plane wave part, we used a Coulomb cutoff of $567\;$eV  ($G_{max} = 6.453$ a.u.); for the augmentation part, we used an $l$-cutoff of $l = 5$. The cutoff for basis envelope functions was $G_\text{cutb}=2.75$ Ry$^{1/2}$ and  cutoff for interstitial part of two-particle screened Coulomb interaction was  $G_\text{cutx}=2.2$ Ry$^{1/2}$. The frequency mesh had a spacing of 0.02 Ry for small $\omega$, the spacing increasing linearly with $\omega$. For calculations with the dipole correction the convergence in the density was approached slowly using the Broyden mixing scheme with a mixing parameter of 0.001 and the Lindhard screening parameter set to zero.
    
    The one-body calculations at the LDA level for the kinetic energy, band structure, and charge density for the finite (bulk) systems were performed with a $28\times 28\times 1$ ($12\times 12\times 3$) $k$-mesh. The relatively smooth dynamical $\Sigma(k)$ was constructed using a $14\times 14\times 1$ ($9\times 9\times 2$) $k$-mesh, and $\Sigma_0 (k)$ was extracted from it, see Fig.~\ref{fig:SF1}. For each iteration in the \qsgw\ self-consistency cycle, the charge density was made self-consistent until the root mean square difference was less than $7\times10^{-7}$ Ry. The \qsgw\ cycle was iterated until $\Sigma_0 (k)$ converged to less than $10^{-5}$ Ry. Thus the calculation was self-consistent in both  $\Sigma_0 (k)$  and the density. Numerous checks were made to verify that the self-consistent  $\Sigma_0 (k)$ was independent of the starting point, for both \qsgw\ and \qsgwh\ calculations, such as using LDA or Hartree-Fock self-energy as the initial self-energy for \qsgw\ and using LDA or \qsgw\ as the initial self-energy for \qsgwh.

\subsubsection{Self-Consistent Ladder BSE, \qsgwh\ Calculations}

\begin{figure}[t] 
    \includegraphics[width=\columnwidth]{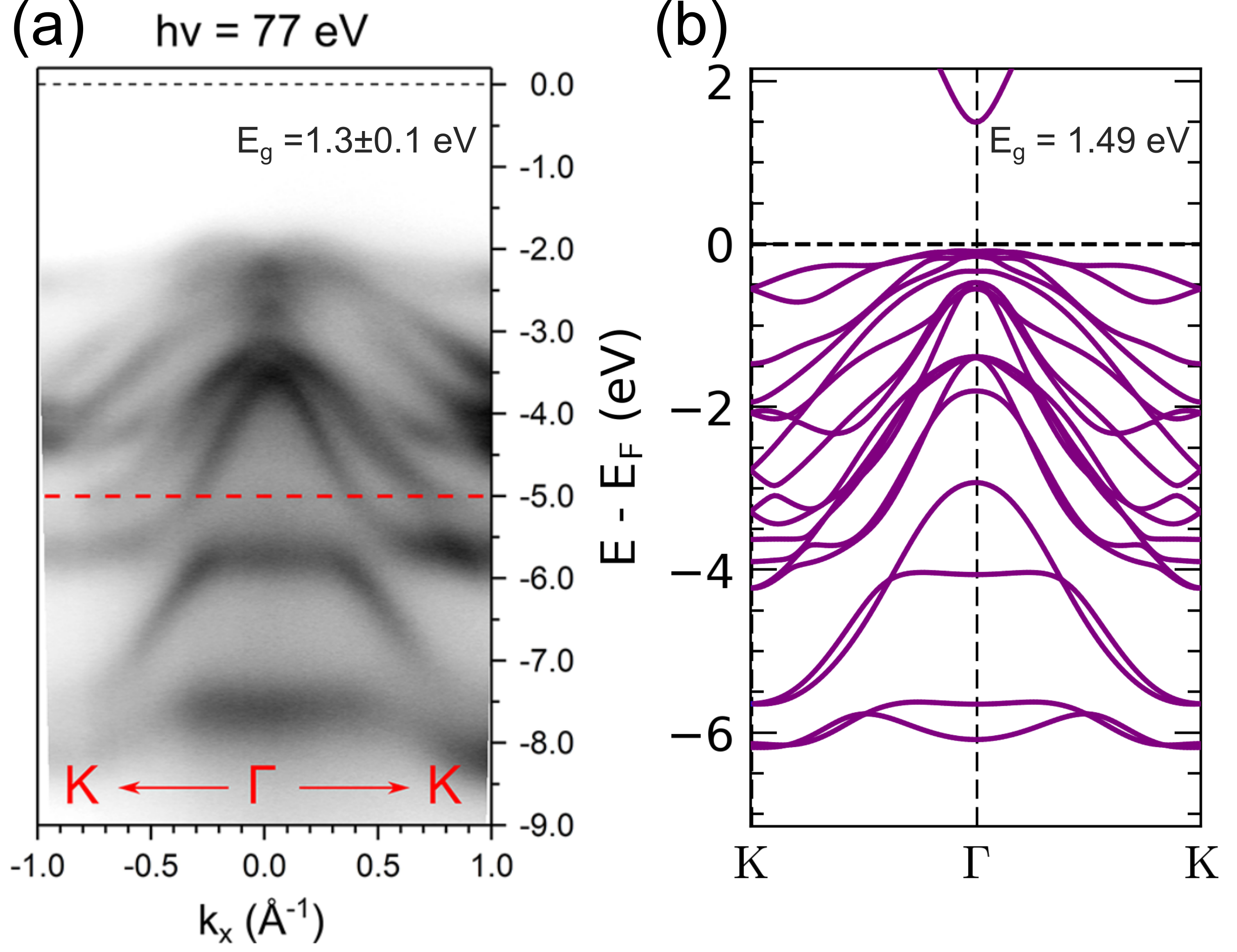}
    \caption{The electronic structure for bulk 2H $\alpha-$\inse\ along the $K-\Gamma-K$ high symmetry path from (a) An energy-momentum cut at 77$\;$eV from a ARPES photon-energy scan with a binding energy of 5$\;$eV (red line). Results\cite{Kremer2023:ACSN} adapted with permission from American Chemical Society Copyright 2023. 
    (b) Using the self-consistent quasiparticle \qsgwh\ approximation by including ladder diagrams. 
    }
    \label{fig:CF2}
\end{figure}

    The electron-hole two-particle correlations are incorporated within a self-consistent ladder-Bethe-Salpeter equation (BSE) implementation\cite{Cunningham2018:PRM,Cunningham2023:PRB} with Tamm-Dancoff approximation.\cite{Hirata1999:CPL,Gruning2009:NL,Rohlfing2000:PRB} The effective interaction $W$ is calculated with ladder-BSE corrections\cite{Bechstedt:2016} and the self-energy, using a static vertex in the BSE. $G$ and $W$ are updated iteratively until both of them converge and this is what we call \qsgwh. Ladder diagram contributions\cite{Bechstedt:2016} increase the screening of $W$, reducing the energy gap besides softening the LDA $\rightarrow$ \qsgw\ corrections noted for the valence bands.
    
    For \inse, we checked the convergence in the \qsgwh\ $E_g$ by increasing the size of the two-particle Hamiltonian. With an increase in the included number of valence and conduction states we observe that the \qsgwh $E_g$ has only minimal changes once 8 valence and 6 conduction states are included in the two-particle Hamiltonian.

    As an example of this computational implementation, in Fig.~\ref{fig:CF2} we show a comparison of the high-resolution angle-resolved photoemission spectroscopy (ARPES) for a bulk $\alpha-$\inse\cite{Kremer2023:ACSN} with our electronic structure calculations using \qsgwh. A very good agreement with the measured results is obtained without any adjustments in the energy gap that would be needed using some DFT approach. More importantly, as we discuss later, this high-level many-body theory description allows us to provide predictive understanding of various FE $\alpha-$\inse~multilayers, without {\em any} adjustable parameters.

\section{Results and Discussion}

\subsection{Structural Considerations}
    \inse~has multiple stable phases that correspond to different arrangements of atoms within a vdW layer.\cite{Kupers2018:IC, Wu2015:NL} We focus on the ferroelectric zinc-blende $\alpha$-\inse~structure depicted in Fig.~\ref{fig:CF1}(a), because it retains stable OOP ferroelectricity down to a single vdW layer and has valuable potential applications.\cite{Zhang2024:NPJCM,Xue2019:AM,Wang2020:AFM} The covalently bound vdW layer is composed of alternating selenium and indium atoms, with the OOP dipole moment arising from the middle selenium atom breaking inversion symmetry by aligning with either indium atom above or below it rather than being between the two. If the selenium atom is below the top indium, the dipole moment points up, while the reverse is true if it is above the bottom indium atom. 

    It has been known that the predicted\cite{Ding2017:NC,Bai2024:ACSN} and measured\cite{Cui2018:NL,Xiao2018:PRL,Bai2024:ACSN,Lv2021:MH,Liu2019:CM}  in-plane component is an order of magnitude greater than the OOP polarization. Several studies have shown that, unlike in other materials, the in-plane and OOP polarizations of $\alpha-$\inse~ are coupled.\cite{Cui2018:NL,Xiao2018:PRL,Lv2021:MH} An applied OOP electric field, not only reverses the OOP polarization but rotates the in-plane polarization and, conversely, an applied in-plane electric field can reverse the OOP polarization. However, the origin and even the existence of a reversible in-plane spontaneous polarization in a single vdW layer of $\alpha-$\inse\ is debated.\cite{Ding2017:NC,Bai2024:ACSN,Xue2019:AM,Wang2020:AFM} As a single vdW layer has a $C_3$ rotational symmetry, it does not meet the symmetry requirements to have ``true'' ferroelectricity in the plane.\cite{Bai2024:ACSN,Pang2025:PRL,Ji2023:Xa} Additionally, under an applied in-plane field, a single domain 1L $\alpha$-\inse\ undergoes an irreversible phase change $\alpha \rightarrow\beta'$,\cite{Bai2024:ACSN} argued to indicate a distinct lack of the in-plane ferroelectricity. However, if defects and  domain wall dynamics are introduced, from the mobility of the middle selenium, there is a reversible in-plane polarization that influences measurable quantities.

    Since the in-plane and OOP polarizations are effectively coupled within a vdW layer,\cite{Cui2018:NL,Xiao2018:PRL,Bai2024:ACSN,Lv2021:MH,Liu2019:CM} in Fig~\ref{fig:CF1}(a) we show a sketch of the combined net polarization in a layer with the diagonal red (blue/black) arrows. The two stacking orders of the vdW layers studied here are 3R, a rhombohedral ABC, and 2H, a hexagonal AB'A stacking. For 2H, the direction of the in-plane polarization alternates between the layers, while for 3R stacking, all the layers have an in-plane polarization in the same direction. The most stable interlayer OOP order in bulk crystals is a FE crystal shown by the middle two structures in Fig.~\ref{fig:CF1}(a), but antiferroelectric (AFE) structures with alternating OOP polarizations are metastable and can be achieved by gating or in films of finite thickness. For simple 2L systems, we show that the character of the band-edge can be tuned by the interlayer OOP polarization of the stacking order.

    In addition to variations of the OOP polarization, we studied the effect of thickness on the band edge. We looked at 2D vdW FE 1L, 2L, and 3L systems. For FE 2L,  we studied both 2H and 3R, while for FE 3L  only the 3R configuration. For the non-FE systems we the studied 3R configuration, either as antiferroelectric (AFE 2L) or ferrielectric (FiE 3L) systems. While there are multiple possibilities for AFE 2L systems utilizing the 2H or 3R configurations, we selected the 3R configuration that uses layers AB and is energetically favorable because it minimizes the surface charges and interlayer repulsion by placing the tetrahedrally coordinated indium atoms near each other.\cite{Kupers2018:IC,Ding2021:ASS,Cui2018:NL} In the non-FE 3L system there is still a net dipole as only two of the OOP dipole moments cancel each other out, which motivates naming it FiE configuration.

    Similar to spintronics, where the OOP magnetization is recognized as key to the scaled-down magnetic memory and efficient magnetization switching, including applications beyond magnetoresistance,\cite{Tsymbal:2019,Zhuravlev2018:APL,Zutic2020:SSC,Dainone2024:N} we focus on the OOP polarization as its control is highly desirable in FEs and their applications.

\subsection{Effects of Out-of-Plane Polarization}
 
    The absence of OOP polarization in vdW systems can arise from the presence of the mirror plane, rotation axis, or inversion symmetry as well, for example, in a 2L configuration if both layers have zero dipole moment. This situation of no OOP polarization corresponds to Fig.~\ref{fig:CF1}(c), where the conduction and valence bands between two identical layers may hybridize as there is an enforced energy degeneracy. In contrast, in Fig.~\ref{fig:CF1}(a) for FE 2H and 3R configurations, there is a broken inversion symmetry, normal to the plane,  and the corresponding OOP polarization. Such a polarization is responsible for the band shift, $\Delta$, as well as for it sign reversal, in the second layer, as depicted in Figs.~\ref{fig:CF1}(d) and~\ref{fig:CF1}(e), showing and important opportunities for FE-controlled systems.

    Here we must distinguish computational artifacts from physical effects. The presence of a surface introduces a depolarizing electric field which reduces the OOP polarization and band shifts. With their finite size in all directions and net charge neutrality, the electrostatic potential of FEs is expected to decay as $1/r^2$ where $r$ is a distance in vacuum away from the FE. In contrast, by using periodic boundary conditions, the periodic slab geometry results in the potential changing linearly in $z$ in the vacuum between the layers, where $z$ is the perpendicular distance from the slab surface. This fictitious long-range interaction between periodically repeated images distorts the band structure.

    This distortion affects both $V_H$ and $V_\text{xc}$. $V_H$ can be corrected with a ``dipole correction'':\cite{Neugebauer1992:PRB} An extra dipole plane is placed in the vacuum, far away from the surfaces, to cancel the artificial linear variation of $V_H$ in vacuum. $V_{xc}$ is more complicated because the nonlocality depends on the approximation used. Common LDA and GGA functionals are local and decay much faster than $1/z$,\cite{Casida2000:JCP} while hybrid functionals, such as HSE06, reintroduce nonlocality through the Fock exchange matrix but include a parameter to set a cutoff distance for the potential. QS$GW$ calculations do not include an arbitrary cutoff distance, thus complicating the effects of artificial periodicity. There is a technique that can deal with this issue,\cite{Tancogne2015:PRB} but we have not implemented it.
    
\begin{figure}[t]
    \includegraphics[width=\columnwidth]{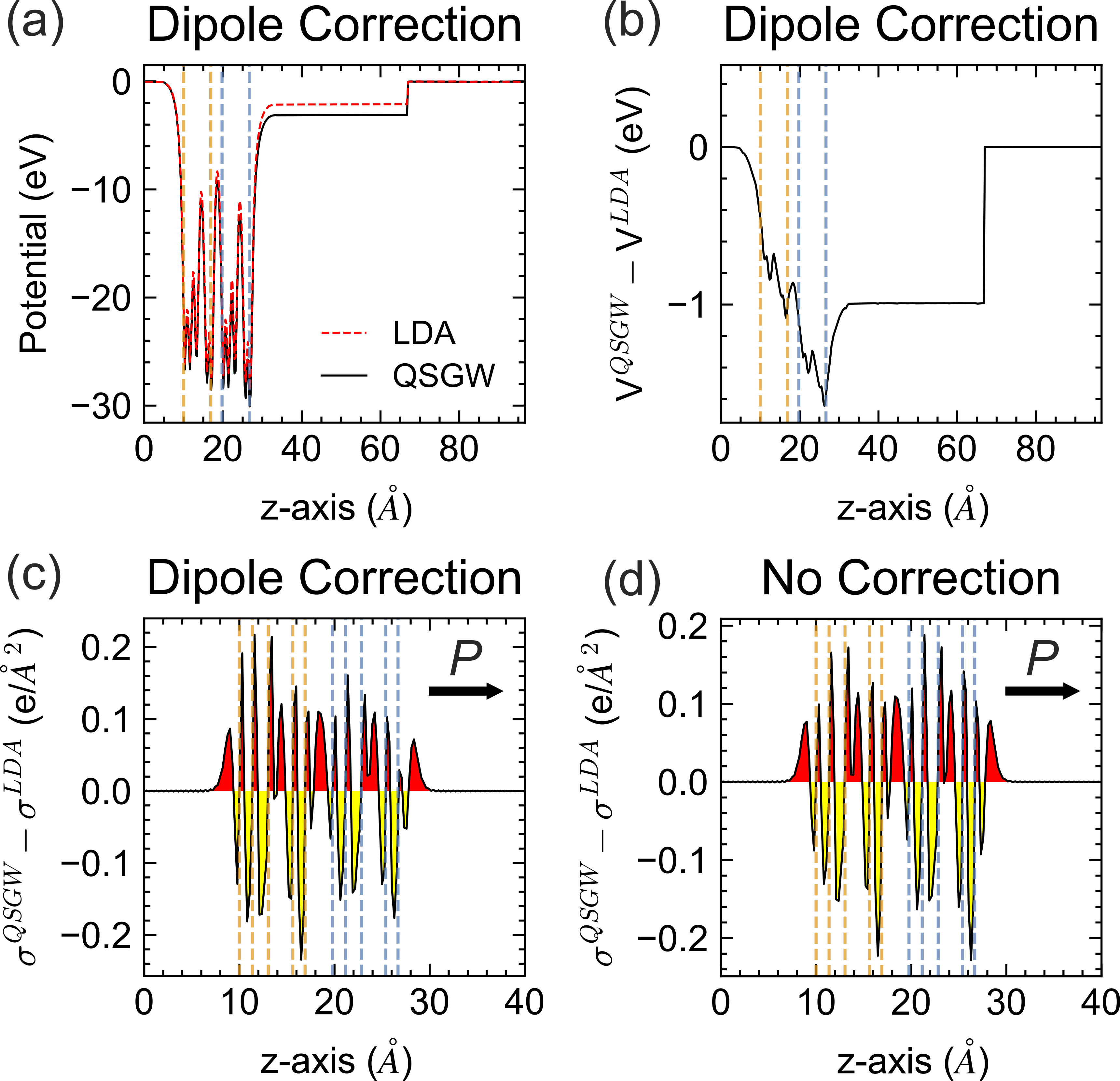}
    \caption{Comparison of LDA and \qsgw\ for FE 3R 2L \inse~ 
    with 80$\;$\AA\
    of vacuum.
    (a)	QS\textit{GW} and LDA dipole corrected averaged 
    planar potentials and (b) their difference, adjusted to vanish at $z=0$. Orange and blue dashed lines: The width of the vdW 2L using the position of the outermost nuclei added to their atomic radii as the edge. 
    (c)	QS\textit{GW} and LDA averaged planar densities with and (d) without the dipole correction. Individually, \qsgw\ and LDA electron densities are both positive. 
    The orange/blue dashed lines: The positions of the atoms within the bottom/top vdW layer. 
    }
    \label{fig:CF3}
\end{figure}
    
    For the QS$GW$ calculations vdW interactions are explicitly included in the nonlocal random phase approximation (RPA) as the second-order bubble diagrams, while higher order diagrams account for other nonlocal effects.\cite{Martin:2016} To correct for the systematic error of inter-image interactions in 2D systems, we used an extrapolation procedure. We held all parameters fixed and recomputed the bands of 3R 2L $\alpha$-\inse\ while varying the vacuum thicknesses (30-100$\;$\AA) and performed a linear extrapolation of the gap value to infinite vacuum thickness. This provides an upper bound for the gap value. 
    
    We have checked the accuracy of this extrapolation procedure by performing calculations with and without the $V_H$ dipole correction, as described in Fig.~\ref{fig:CF3}, while the extrapolated gap values were within 0.01$\;$eV ($\sim$1\%) of each other [Figs.~\ref{fig:SF2}(b) and ~\ref{fig:SF2}(c)]. As a result, the dipole correction was only used in Figs.~\ref{fig:CF3} and \ref{fig:CF4}, as it increased the computational expense at the QS$GW$ level by requiring a slow mixing scheme for the density that is prone to falling into local minima. To the best of our knowledge, our results are the first implementation of the dipole correction in a self-consistent many-body approach.

    The potential step in the vacuum region of Fig.~\ref{fig:CF3}(a) arises from the dipole correction applied to the FE 3R 2L potential, shown for both the LDA (red) and QS$GW$ (black) calculations. The vertical dashed lines schematically delineate the bottom and top surfaces of the two vdW layers. The resulting change in electrostatic potential is 2.107$\;$eV for LDA and 3.044$\;$eV for QS$GW$. This potential drop is proportional to the OOP dipole moment per unit area of the system, which for FE 3R 2L $\alpha$-\inse\ is $1.88\times 10^{-11}\;$C/m at the LDA level and $2.76\times 10^{-11}\;$C/m at the \qsgw\ level. When compared to previously reported HSE06 (GGA) dipole moments, this \qsgw\ moment for the 1L (2L) system is $\sim$25\% ($\sim$50\%) larger.\cite{Ding2017:NC} By subtracting the LDA potential from the QS$GW$ potential in Fig.~\ref{fig:CF3}(b) it is easier to see that the potential around the atomic sites is also different for the two approximations.  

    Surprisingly, this suggests that in systems with large OOP higher-order theories beyond DFT are required, even to describe {\em ground-state properties}, such as the full dipole moment. In both DFT and hybrid functionals, the converged density is built upon an approximate and fixed $V_\text{xc}$, which produces an overscreened density in insulators. Consistently with this, DFT predicts a metallic state and HSE06 yields a reduced band gap, as seen in Figs.~\ref{fig:CF4}(a)-\ref{fig:CF4}(c). The resulting small excitation energies imply an artificially large polarizability [recall Eq.~(\ref{eq:RPA_Polarizability})], which suppresses the magnitude of the dipole moment. In contrast, the self-consistently updated $\Sigma$ improves the treatment of the exchange and correlation contributions, reducing spurious screening and correcting the dipole moment. Ensuring that the dipole moment is accurate is critical to making reasonable predictions for heterostructure devices utilizing \inse, as the key properties often rely on the field produced by the OOP polarization.\cite{Bai2024:ACSN,Si2019:NE,Huang2022:IM,Ahn2006:RMP} Additionally, this potential drop across the film thickness also means the two surfaces of the slab have different work functions, a feature that has been used for catalysis in other FEs.\cite{Zhang2012:CR}

    We have also examined the influence of the dipole correction in and the LDA and QS$GW$ for the averaged planar electronic charge densities, shown in Figs.~\ref{fig:CF3}(c) and ~\ref{fig:CF3}(d), for their corresponding difference. The vertical lines indicate the location of the atoms within a vdW layer, while the black arrow indicates the OOP polarization (the direction of the net dipole moment). There is a pronounced charge accumulation (depletion) in red (yellow) to the right (left) of each atomic site because the dipole moment of the film is larger in QS$GW$. In the next section we will show that the large deviations between the LDA and \qsgw\ can be explained by the fictitious eigenvalues of the DFT Kohn-Sham equations discussed in Sec. IIA. As the LDA polarizability of the system depends on these one-particle eigenvalues, their inadequacy results in an incorrect polarizability and a predicted metallic gap. Consequently, the LDA results suggest an artificially increased screening which reduces the electric field within the material and the resulting strength of the OOP dipole. In contrast, the QS$GW$ results correctly identify the insulating \inse~ which lead to a reduced screening and a larger displacement of the charges within the material.

\subsection{Electronic Structure of In$_2$Se$_3$}
\subsubsection{Effect of Different Approximations}
   
    While our preceding results clearly show pronounced differences between using the LDA and QS$GW$ approach, it is important to better understand if this difference is simply an isolated case and other DFT approximations could provide a better description, or the problem with using DFT for 2D vdW FEs is more serious. Furthermore, this is also an opportunity to examine the relevance of different $GW$ implementations as they are inequivalent approximations and do not correspond to the same fidelity. Some guidance what to expect is provided by recalling known experimental findings.

\begin{figure*}
    \centering
    \includegraphics{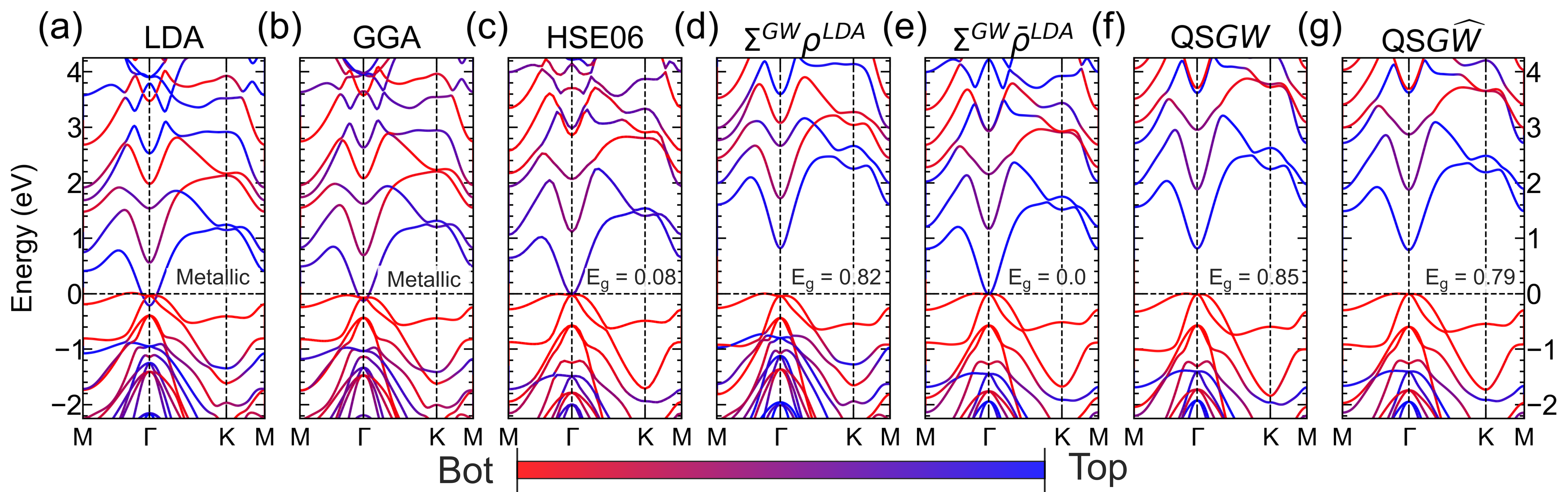}
    \caption{Electronic  structure and the global gap, $E_g\;$(eV) 
    of FE 3R 2L $\alpha$-\inse\ for 
    different approximations, 
    computed with the general dipole correction and 80$\;$\AA~of vacuum. The polarization of both layers is pointing up. The top (bottom) vdW layer is denoted by blue (red).  
    (a) Local density approximation (LDA), (b) PBE generalized gradient approximation (GGA), (c) Hybrid functional HSE06, (d) Modified single-shot 
    GW with the LDA charge density, 
    $\rho^\text{LDA}$, and the quasiparticle self-energy, $\Sigma^{GW}$. 
    (e) Single-shot GW with 
    the updated charge density, $\bar \rho^\text{LDA}$ and the quasiparticle self-energy, with the same modifications. 
    (f) Quasiparticle self-consistent GW (\qsgw), and (g) \qsgw\ with Bethe-Salpeter equation vertex correction (\qsgwh).}
    \label{fig:CF4}
\end{figure*}

    A number of studies have experimentally explored $E_g$  of $\alpha$-\inse~[recall from Fig.~\ref{fig:CF1}(b) different types of band gaps] with $E_g^\text{bulk}=1.17 - 1.45\;$eV depending on the stacking structure and measurement technique.\cite{Kremer2023:ACSN,Ho2013:ACSAMI,Quereda2016:AOM,Felton2025:NC,Lyu2020:NT,Zhang2022:JPCL,Cho2022:PRM,Qasrawi2006:TSF,Ye1998:JJAP} It was also shown that $E_g$ is thickness-dependent, increasing from 1.45 to $2.5 \pm 0.3\;$eV.\cite{Quereda2016:AOM,Lyu2020:NT,Zhang2022:JPCL,Cho2022:PRM} In these studies $E_g$ gap increases in films thinner than 20$\;$nm, with the largest optical gap for single vdW layer flakes grown by vapor deposition. On the computational side, DFT has been used for the bulk,\cite{Kremer2023:ACSN,Quereda2016:AOM,Kremer2023:ACSN,Felton2025:NC} multi-layered,\cite{Ding2017:NC,Ding2021:ASS,Debbichi2015:JPCL,Huang2020:PRB,Lyu2020:NT} and 1L\cite{Ding2017:NC,Ding2021:ASS,Huang2020:PRB,Hu2017:RSCA,Bai2024:ACSN} systems. While $E_g$ is systematically underestimated in the bulk and 1L systems, for systems with two or more layers DFT makes the qualitative prediction that $E_g=0$. 

    Our computed results in Fig.~\ref{fig:CF4} examine the band structure of a 3R 2L system for different approximations. In Figs.~\ref{fig:CF4}(a) and \ref{fig:CF4}(b) we first consider the commonly used LDA and GGA functionals, both showing that $E_g$ is closed. There has been reasonable success in higher-order corrections increasing $E_g$ for the 1L and bulk systems,\cite{Ding2017:NC,Hu2017:RSCA,Huang2020:PRB,Quereda2016:AOM} and even opening $E_g$ for 2L systems.\cite{Hu2017:RSCA,Huang2020:PRB,Ding2021:ASS} The studies that did consider the few-layer limit concluded that reintroduction of nonlocal exchange and correlation effects are necessary to reopen $E_g$, at least for the 2L system.\cite{Hu2017:RSCA,Huang2020:PRB,Ding2021:ASS} 

    However, for methods that utilize Hubbard parameters and hybrid functionals, $E_g$ is matched to experiment by tuning parameters that are unique to a given configuration, thus are not predictive when applied to heterostructures, nor true first-principles methods. For our parameterized Hamiltonian, we chose the hybrid functional HSE06 in Fig.~\ref{fig:CF4}(c), as the inclusion of the Fock exchange matrix and an additional screening parameter is known to overcome many of DFT's limitations. For illustrative purposes, we did not arbitrarily tune the mixing parameter $\alpha$ in Eq.~(\ref{eq:hybrid}) to open the gap and instead use the baseline $\alpha=25$\% of the nonlocal Fock exchange recommended by VASP, resulting in a vanishingly small $E_g=0.08\;$eV ($E_g=0.0\;$eV in the prior version). As the electrostatic potential drop is comparable to the untuned enlarged $E_g$, the electron transfer between layers yields an effectively metallic self-consistent solution. However, by increasing $\alpha$ or the screening, $E_g$ can be enlarged.

    Turning next to the $GW$ description it is important to recognize different implementations and their fidelity. For \inse,~\textit{GW} approximation has been limited to $G^\mathrm{PBE}W^\mathrm{PBE}$ for bulk\cite{Debbichi2015:JPCL,Quereda2016:AOM} and 1L systems,\cite{Debbichi2015:JPCL,Fang2023:PRA} while the  2L calculation\cite{Cho2022:PRM} was only realized for $\Gamma$-point. Computed bulk results were in good agreement with experiments to yield $E_g= 1.29{-}1.4\;$eV, while for 1L $E_g=1.92\;$eV was an underestimate. Similar to our discussion from Sec.~II, an important limitation of such ``single-shot" $G_0W_0$ calculations is that the result depends on the starting point of the independent-particle description.\cite{Ismail2017:JPCM,Schilfgaarde2006:PRB,Cunningham2023:PRB} While we did not revisit the bulk and 1L systems with $G_0W_0$, we used a modified single-shot method to compute the 3R 2L system, as shown in Fig.~\ref{fig:CF4}(d). We used $G^\mathrm{LDA}W^\mathrm{LDA}$ which differs from the usual $G_0W_0$ counterpart in two respects: The off-diagonal parts of the (quasiparticlized) self-energy $\Sigma_0$ were also included, and we set the renormalization factor~$\textit{Z=1}$, which can still provide an estimate for a limited type of self-consistency, as noted in the Appendix of a previous work.\cite{Schilfgaarde2006:PRB} Generally, $E_g \propto 1/Z$. As it is typically done with $G_0W_0$, we kept  $n(\mathbf{r})$ unchanged when computing the quasiparticle band structure. As shown in Fig.~\ref{fig:CF4}(d), we obtain $E_g >0$ under these conditions.

    However, the gap opening seems to be a fortuitous accident for $\alpha$-\inse. Strikingly, $E_g$ closes again if the density $n(\mathbf{r})\equiv \rho^\text{LDA}$ is not fixed and allowed to be self-consistently updated using the $\Sigma_0$ acquired from $G^\text{LDA}W^\text{LDA}$, see Fig.~\ref{fig:CF4}(e). In this case, the off-diagonal elements of $\Sigma_0$ modify the density, $\bar \rho^\text{LDA}(\mathbf{r})$, and the corresponding $V_{xc}(\mathbf{r})$. This potential change, $\Delta{V}$, can be estimated if we start from the assumption that the inverse linear polarizability, $\chi^{-1} = \delta{V}/\delta{\rho^\text{LDA}}$, acquired from LDA is adequate. Thus, when $\rho^\text{LDA}$ is modified to $\bar \rho^\text{LDA}$, the potential under this assumption becomes 
    \begin{gather}
        V(\bar \rho^\text{LDA}) = \Sigma_0 - V^\mathrm{LDA}(\rho^\mathrm{LDA}) + V^\mathrm{LDA}(\bar \rho^\text{LDA}), 
        \label{eq:Vbar}
    \end{gather}
    where the charge modification is driven by the self-consistent addition of a fixed external potential $\Sigma_0 - V_\text{xc}^\mathrm{LDA}(\rho^\mathrm{LDA})$ to the LDA Hamiltonian. A similar situation applies to TiSe$_2$, where $G^\mathrm{LDA}W^\mathrm{LDA}$ makes $E_g$ positive,\cite{Cazzaniga2012:PRB,Pashov2025:NPJCM,Acharya2021:NPJCM} but making $n(\mathbf{r})$ self-consistent in both DFT and \qsgw\ makes the indirect gap negative again. 

    For $\alpha$-\inse, \qsgw\ yields $E_g>0$ as seen in Fig.~\ref{fig:CF4}(f), but for very different reasons than within the $G^\mathrm{LDA}W^\mathrm{LDA}$. The single-shot gap opening arises from a exchange driven correction on a static metallic density, but is unable to alter this ground state.\cite{Caruso2014:PRB} Therefore, it gets the right answer for the wrong reason. In contrast, \qsgw\ iteratively updates both the density and the self-energy.\cite{Kotani2007:PRB} This mutual self-consistency lets the nonlocal exchange interaction increase to compensate for  the  reduced screening of $\bar \rho$ from the decreased polarizability when the gap opens. In analogy with  hybrid functional approaches, \qsgw\ determines an optimally tuned amount of nonlocal exchange, but with a fully {\em ab initio}, $\mathbf{k}$- and orbital-dependent ``mixing'' determined by the system’s own screening, rather than a single scalar mixing parameter $\alpha$ as in HSE06. The closing of the gap when $n(\mathbf{r})$ was updated after introducing the non-local exchange in Fig.~\ref{fig:CF4}(e), shows that this initial $n(\mathbf{r})$ is incorrect and that \qsgw\ is needed to find the proper ground state $n(\mathbf{r})$ by rotating the one-body basis towards the quasiparticle one.

    In Fig.~\ref{fig:CF4}(g) BSE was used as a vertex correction to improve the screened Coulomb interaction ($\widehat{W}$)  by including the ladder diagrams to account for electron-hole interactions. In strongly-correlated systems the correction to $E_g$ can be dramatic, on the order of several eVs. However, being a weakly correlated $s-p$ system, the correction is 0.06$\;$eV showing that this much more expensive correction is not needed. As implemented in Questaal and other codes starting at a Bloch like theory, RPA treats long range interactions reasonably well. Thus the corrective effect of the ladder diagrams is most stark when RPA underestimates the local screening of the exchange interaction.\cite{Cunningham2023:PRB} In particular, systems with localized $d$ states or flat bands are likely to respond more strongly to the vertex correction. 
 
    The use of ladder diagrams in other 2D FE systems would involve a few considerations. First, the band gaps in light-mass polar materials are systematically overestimated.\cite{Cunningham2023:PRB} This is because in those systems, and in many FEs, the electron-phonon interaction plays a substantial role in the renormalization of $\Sigma$.\cite{Hedin1965:PR} Additionally, due to the omission of the exact vertex ($\Gamma$ in Hedin's equations)\cite{Hedin1965:PR,Martin:2016} in the exact self-energy, dispersionless core-like $d$ states at the Fermi level are not pushed down enough, thus the predicted gap will be underestimated in these cases.\cite{Cunningham2023:PRB} This is an issue of the vertex correction not including \emph{all} diagrams, only the ladder diagrams.\cite{Cunningham2023:PRB} Finally, there is the question of what quantity is being studied. Beyond the band gap renormalization, the inclusion of the ladder-diagram vertex correction also provides an improved dielectric response, screening, and quantified optical properties such as oscillator strengths and excitons. Taken together, if these other quantities are directly considered in a study, the added computational expense from the vertex correction is justified. With this in mind, and if considering only the band gap renormalization, we expect Moir\'e FEs with flat bands or sliding FEs comprised of 2L transition metal dichalcogenides (e.g., MoS$_2$, WS$_2$, MoSe$_2$, and WSe$_2$), would be most affected by the vertex correction.\cite{Wu2021:PNAS} Specifically, the addition of the vertex correction in 1L WSe$_2$ decreases the band gap by $\sim0.25$\;eV, while the transition from LDA to QSGW adjusts the gap by $\sim1.4$\;eV from 1.53\;eV to 2.95\;eV.\cite{Dadkhah2024:PRB,Thodika2025:X}

\begin{figure*}
    \centering
    \includegraphics{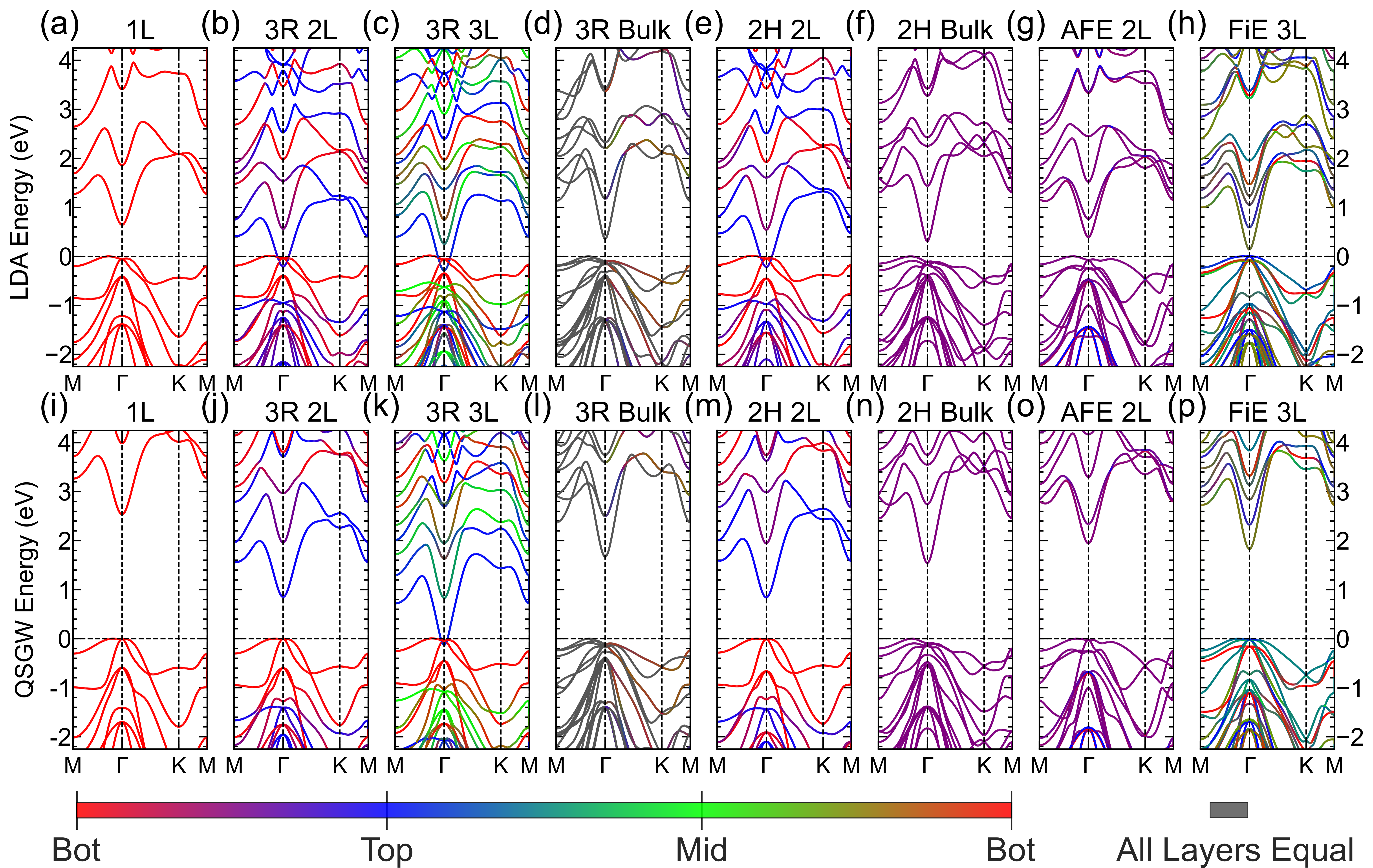}
    \caption{Comparing the LDA and QS$GW$ electronic structure of different configurations. 
    (a)-(d) LDA and (i)-(l) QS$GW$ 
    for the 3R $\alpha$-\inse\ configuration. The first three columns are for 1L, 2L, and 3L,
    while the last is for the bulk. (e), (m) 2L and  (f), (n) bulk electronic structure for the 2H $\alpha$-\inse\ configuration. (g), (o) AFE 2L and (h), (p) FiE 3L 3R
    $\alpha$-\inse.}
    \label{fig:CF5}
\end{figure*}  
   
\subsubsection{Bands in a Capacitor Model} 

    While a complete description of ferroelectricity requires quantum mechanics, our semiclassical model of a band structure in a capacitor from Fig.~\ref{fig:CF1}(b) already provides valuable insights. With the band bending that is slow at the atomic scale, this model shows a useful distinction between the local, $E_{g}^\mathit{local}$, and global band gap, $E_g$. Here $E_{g}^\mathit{local}$, measures the energy difference between the conduction and valence band edges at a given point in space.  $E_g$ is the difference between the lowest conduction and highest valence band energies across the entire thickness of the capacitor; the latter is highlighted in Fig.~\ref{fig:CF1}(b) by the horizontal dashed line. In the absence of the depolarizing field and surface effects, the band edges would lack spatial dependence and would be a pair of horizontal lines. In this limit,  $E_g=E_{g}^\mathit{local}=E_g^\mathit{bulk}$. If we now introduce the capacitor the potential drop across its thickness introduces a spatially dependent electrostatic shift, $\Delta$, in the energy of the bands causing the bands to tilt (dashed colored lines). In this case, $E_g=E_g^\mathit{local}-\Delta$. As $\Delta$ gets larger, either due to a larger magnitude of charge for a fixed thickness or a sufficiently thick capacitor, $E_g$ will decrease until the band edges meet, resulting in no global gap. In addition to the tilt, if the capacitor is sufficiently thin, quantum confinement can substantially increase the value of $E_g^\mathit{local}$ near the surfaces of the film.

    This capacitor model translates well to the FE film calculations. Here the plates of the capacitor are the surfaces of the FE, as the OOP polarization introduces a potential drop across the thickness of the dielectric material that results in the accumulation of charges on the surface. In ambient experiments, these bare surface charges would usually be strongly compensated due to surface reconstruction, FE domains or some other material being electrostatically bound to the surface. However, for high quality samples cleaved under ultra-high vacuum ($<5\times 10^{-9}$ mbar), high-resolution ARPES have observed highly metallic 2D electron gases (2DEGs) confined to the surface.\cite{Kremer2023:ACSN} In their interpretation, surface defects give rise to the existence of the 2DEG in \inse, as 2DEGs have been observed at the surface of other FE systems or heterojunction interfaces where oxygen vacancies played a key role in doping the system.\cite{Brehin2020:PRM,Wang2018:ACSAMI,Yu2014:NC}

    In this defect-doping picture, positively charged chalcogenide vacancies induce band bending at the surface of the material.\cite{Cook2019:AEM} This provides a confining potential well for the 2DEG, which shifts the chemical potential and creates a charge accumulation layer near the surface.\cite{Monch:1993,Cook2019:AEM} In that regime, the measured surface metallicity arises from  chalcogenide vacancies and Fermi-level pinning rather than by polarization-driven electronic reconstruction.\cite{Cook2019:AEM} This mechanism is particularly relevant in cases where samples are vacuum annealed as the low-pressure heat treatment provides suitable conditions for the formation of chalcogenides vacancies near the surface.\cite{Zhu2017:ACSN} The capacitor model elucidates a complementary mechanism that coexists with the established defect-induced doping but is most relevant for defect-free flakes in a controlled vacuum environment, under the assumption of weak screening. In thicker \inse\ flakes, typically on the order of tens of nanometers, the depolarizing field energy is reduced via the formation of OOP domains; for example, a bi-domain structure was observed in a 35\;nm thick flake.\cite{Masood2025:JAP} This suppresses the net macroscopic potential drop when averaged over many domains, thus changing the global gap. In contrast, for sufficiently thin films, roughly two vdW layers ($\sim1.6$\;nm) up to $\sim$10--20\;nm, the semiclassical layer-resolved picture is meaningful.

     In the thin-film weakly screened regime, our calculations predict that $E_g$ closes beyond approximately 3 vdW layers, consistent with the emergence of metallic 2DEG surface accumulation discussed above. Regardless of the theoretical approximation used there will always be a film thickness resulting in $E_g=0$, due to the electrostatic shift between vdW layers. This is reflected in our calculations in Fig.~\ref{fig:CF5}: Each vdW layer is distinguishable by the orbital contributions to the bands, displaying interlayer electrostatic band bending. Red (blue) denotes bottom (top) layer, while in 3L systems the middle layer is green. Hybridization between layers is indicated by a mixing of colors. In 2L systems, fully hybridized bands are purple, while in 3L they are gray. For all non-bulk systems a vacuum of 80$\;$\AA\ was used.
     
    For different DFT approximations, known to systematically underestimate the gap, in Figs.~\ref{fig:CF5}(a)-\ref{fig:CF5}(c) we find $E_g=0$, already in 2L systems. With corrections such as HSE06, DFT+U, or $G_0W_0$, $E_g$ can be opened for a few layers but $E_g=0$, if enough layers are considered, as shown in the Fig.~\ref{fig:CF5}(k) for QS$GW$ in the 3L case. While systematically making a metallic prediction for many-layered systems may seem to obviate the need for QS$GW$ for this type of system, the analysis of $E_g^\mathit{local}$ for each layer shows that QS$GW$ aligns with experiments across all thicknesses, with any deviations being attributable to systematic errors. With $E_g=0$, the magnitude of the dipole moment will saturate, as charge carriers will tunnel from one side of the slab to the other. As a result, the ``slope'' of the potential drop across the thickness of the material begins to flatten out when more vdW layers are added. Thus, taken to the near-bulk limit, any individual vdW layer would effectively experience $\Delta \approx 0$, reintroducing the symmetry broken by the field in the thin films resulting in an energy degeneracy. In this regime, the hybridization between layers would be stronger, as the semiclassical picture breaks down and individual layers are no longer a ``good'' quantum number. Instead, the hybridization returns to Bloch states where eigenstates, $k_z$, are superpositions over all layers in the bulk. 

\begin{table}
\begin{tabular}{ccccc}
    \hline \hline
    Units (eV) &  \multicolumn{2}{c}{$E_g$} & \multicolumn{2}{c}{Average $E_g^{local}$}  \\
    &  LDA  & \qsgw &  LDA   & \qsgw \\ 
    \hline
    FE 1L & 0.64 & $2.54\pm0.25$  &  -  &  - \\
    FE 2H 2L & 0.0 & $0.84\pm0.25$  & 0.6  & 2.17 $\pm0.25$ \\
    FE 3R 2L & 0.0 & 0.85 ($1.1^*$) & 0.6 & 2.18 $\pm0.25$ \\
    FE 3R 3L & 0.0 & 0.0 & 0.81 & 1.91 \\
    \hline
    2H Bulk & 0.31 & 1.54 &  -  &  -  \\
    3R Bulk & 0.35 & 1.67  &  -  &  -  \\
    AFE 3R  2L & 0.39 & 1.94  &  -  &  -  \\
    FiE 3R 3L & 0.14 & 1.83$\pm0.25$    &  -  &  -  \\
    \hline \hline
\end{tabular}
    \caption{The LDA and QS$GW$ gaps for $\alpha-$\inse.~
    An asterisk marks the extrapolation to infinite vacuum thickness. Without this extrapolation 
    there is a systematic underestimation 
    $\sim 0.25\;$eV from the ``true''  QS$GW$ value for few-layer systems with an OOP polarization. The average local band gap was determined for each layer and then averaged.}
    \label{tab:1}
\end{table}    
     
\subsubsection{Ferroelectric Configurations}

    To complement our findings from Figs.~\ref{fig:CF4} and \ref{fig:CF5} and the intuition from the capacitor model [see Fig.~\ref{fig:CF1}(b)], we summarize the LDA and \qsgw\ global and averaged local gaps gaps in Table~\ref{tab:1} for the considered configurations. We first focus on the FE systems. In the bulk, the LDA underestimates $E_g$, but both for 2H and 3R configurations $E_g>0$, while the for \qsgw, as expected, $E_g$ is overestimated.\cite{Kotani2007:PRB} This \qsgw\ issue is addressed by including the ladder diagrams and corrected polarizability in \qsgwh, where for 2H configuration $E_g=1.49\;$eV, Fig.~\ref{fig:CF2}(b) is within 3\% of optical measurements\cite{Quereda2016:AOM,Lyu2020:NT,Ho2013:ACSAMI} and 7\% of the ARPES.\cite{Kremer2023:ACSN}

    For 1L, 2L, and 3R results, an overestimate in the \qsgw\ gap is offset by its underestimate from the periodic boundary conditions. To address this issue, we computed the gap at several vacuum thicknesses and extrapolated to infinite thickness, as shown in Fig.~S\ref{fig:SF2}. The explicit dipole correction was only applied for the 2L 3R system, as the computational expense scaled poorly with the vacuum thickness. Therefore, we limited the use of \qsgwh\ to the 2L and bulk calculations. With this extrapolation procedure, for 3R 2L system the \qsgw\ $E_g$ increased from 0.85 to 1.1$\;$eV, indicating that the systematic underestimate of $E_g$ is on the order of $0.25\;$eV.
    
    For \qsgw in FE 3L, we find $E_g=0$, but with 80$\;$\AA\ of vacuum $E_g^{local}=1.91\;$eV. This system is also influenced by fictitious inter-image interactions so $E_g^{local}$ increases with vacuum thickness. However, the systematic error determined by the vacuum extrapolation for the 2L system cannot be applied to the metallic 3L. The computed local gaps for 1L-3L are 1.9-2.5$\;$eV, within the range for measured films with several layers.\cite{Quereda2016:AOM,Lyu2020:NT,Zhang2022:JPCL,Cho2022:PRM} Tracking the conduction band edge at the $\Gamma$-point for the bottom layer (red) across Figs.~\ref{fig:CF5}(i)-~\ref{fig:CF5}(k), at $\sim 1.7\;$eV in 3R 3L, we see that, as each vdW layer is added, the band hybridizes with other layers and monotonically approaches the bulk gap, as expected from the capacitor model.

    Considering that the rotation of vdW layers can have a substantial effect on the measured properties of 2L systems, we have also computed the 2L  and bulk 2H configurations, where one of the \inse\ layers is rotated by 60$^\circ$ relative to the 3R system, which also changes the relative alignment of the in-plane polarization between the layers. For 2L, in Figs.~\ref{fig:CF5}(e) and \ref{fig:CF5}(m), the difference  between the 2H and 3R systems was minimal for both the gap and the band dispersions. For the bulk, in  Figs.~\ref{fig:CF5}(f) and \ref{fig:CF5}(n), we find that the change in the gap is relatively small, $\sim 10\;$meV, but the hybridization and dispersions of the bands, particularly along the $\Gamma-K$ path, change dramatically. This can be partially attributed to the two different stacking orders having different symmetries. 

\subsubsection{Nonferroelectric Configurations}
    
    For the non-FE systems we examined only those with a few layers, given that the bulk system is FE. In both AFE and FiE systems for the LDA there is $E_g>0$ that is smaller for FiE, consistent with the corresponding decrease in the net OOP polarization. Without any OOP polarization for the AFE system, considered in Figs.~\ref{fig:CF5}(g) and \ref{fig:CF5}(o), there are energetically degenerate states near the band edge, as expected from Fig.~\ref{fig:CF1}(c). There is also an interlayer hybridization, absent in the FE  thin films, seen from the change in the band dispersion along the $\Gamma-K$ high-symmetry line when compared to the FE 1L, 2L, and bulk results from Figs.~\ref{fig:CF5}(i), \ref{fig:CF5}(j), and ~\ref{fig:CF5}(l). While 1L and 2L 3R configurations have nearly identical valence band edges with the conduction band edges also being similar and shifted in energy, in the bulk the band edge is altered as there is a interlayer hybridization from Bloch states normal to the vdW plane.\cite{Kremer2023:ACSN,Martin:2016} The AFE 2L band edge is unlike the one in FE thin films or the bulk,  indicating that there is an  interlayer hybridization but it arises from a different mechanism than in the bulk as for 2L there are no allowable Bloch states.
    
    In FiE 3L system, shown for the LDA and \qsgw in Figs.~\ref{fig:CF5}(h) and ~\ref{fig:CF5}(p), there is a partially compensated  OOP (up-down-up) polarization. Just as the polarization magnitude, the observed FiE behavior is between the AFE and FE counterparts. The net OOP polarization removes the energy degeneracy seen from the AFE 2L configuration but, unlike the FE thin films, the states are still close enough in energy to hybridize. LDA and QS$GW$ predict different hybridization at the valence band edge. In the conduction and valence edges, the middle layer (green) hybridizes with the other two layers, but there is minimal hybridization between the top and bottom layers. Along the $M-\Gamma$ path the LDA bands show three distinct layers with minimal hybridization with dispersions from each layer matching the 1L case. For QS$GW$, the top and middle layers hybridize while the bottom layer matches the 1L case. The conduction band edge at both approximations match closely, however, now the hybridization is between the bottom and middle layers.

\section{Conclusion and Outlook}

    By focusing on a two-dimensional ferroelectric $\alpha$-In$_2$Se$_3$, we show that the out-of-plane polarization is responsible for strong modifications in the electronic structure which requires a higher-level first-principles description, beyond density functional theory. This is surprising for the description of the considered ground state properties, such as the calculated charge densities, which should be well represented even with the low-level theories. Instead, our results show that not only the common local functionals and more sophisticated hybrid functionals are insufficient, but even the non-self-consistent {\it GW} approaches may fail to provide an adequate description. 

    In a monolayer or bulk $\alpha$-In$_2$Se$_3$ one can argue that the known underestimated values of the energy gap, calculated  with common first-principles methods, can be corrected by the comparison with the experimental results. However, this approach is no longer feasible in multilayer structures where other implications, such as the band bending or hybridization between different bands cannot be simply adjusted by a scissors operator, which only introduces a rigid shift of the calculated bands to match the value of the experimental global gap. Even in prescriptions where such a shift produces the correct gap (by adding $U$ to density functional theory), adjusting one quantity to fit experiment, such as the band gap, does not ensure that other physically important ones are well described.  This is also seen from the ambiguities in one-shot \textit{GW} results. For multilayer ferroelectrics, we argue that using an accurate implementation of the self-consistent quasiparticle {\it GW}, as available in a versatile open-source package Questaal,\cite{Pashov2020:CPC} becomes indispensable. This is supported by exploring different ordering in a single material, becoming ferroelectric, antiferroelectric, or ferrielectric, depending on the stacking configuration or the number of layers. The corresponding changes in the strength of the out-of-plane polarization directly influence the scale of deviation of the density functional theory from the more accurate many-body description.

    Our results, focused on $\alpha$-In$_2$Se$_3$, suggest important opportunities for future studies that could also involve other two-dimensional ferroelectrics. For example, since we found (Fig.~\ref{fig:CF3}) that the out-of-plane polarization from the self-consistent quasiparticle {\it GW} is about 50$\;$\% larger than from the density functional theory for bilayer ferroelectric, it would be helpful to revisit prior studies where a similar many-body description was not employed and the role of ferroelectricity may have been underestimated. This is also relevant in the studies of proximity effects, where, for many decades, both magnetic and ferroelectric effects were only analyzed within the single-particle picture,\cite{Yeh1968:PSS,Hauser1969:PR} but many-body effects and accurate description of correlation effects could be essential.\cite{Scharf2017:PRL,Zutic2019:MT,Deb2024:NC,Zollner2020:PRB,Xu2020:PRL} An implementation of the self-consistent quasiparticle {\it GW} and Bethe-Salpeter equation in Questaal  shown provides a suitable computational framework to explore many-body manifestations of such proximity effects.\cite{Thodika2025:X}

    A large out-of-plane polarization could offer an effective ferroelectric control of topological states.\cite{Zhang2021:NL} An intrinsic magnetic topological insulator, MnBi$_2$Se$_4$,\cite{Zhu2021:NL} with its in-plane magnetization, is not expected to support quantum anomalous Hall effect, unlike a much better studied MnBi$_2$Te$_4$\cite{Chang2023:RMP} with an out-of-plane magnetization. However, through ferroelectric proximity effect to change the magnetic anisotropy, could this situation be altered and an out-of-plane magnetization, compatible with the quantum anomalous Hall effect, be also realized in MnBi$_2$Se$_4$? Similarly, realizing quantum spin-valley Hall kink states which, with spin-valley-momentum locking, are robust against different types of disorder\cite{Zhou2021:PRL} could be simplified using a ferroelectric control. Rather that employing a complex fabrication of planar heterostructure including two materials that separately support quantum spin Hall and quantum valley Hall states,\cite{Zhou2021:PRL} which are distinguished by the relative strengths of  spin-orbit coupling, a single material could suffice such that the needed change in the strength of the spin-orbit coupling would be realized by its partial covering with a two-dimensional ferroelectric.

    Despite our focus on the implementation of many-body description within the open-source package Questaal,\cite{Pashov2020:CPC} we expect that described hierarchy of different first-principles methods  will also motive efforts to consider the advantages of the self-consistent quasiparticle {\it GW} approximation in other computational frameworks. 

\section*{Acknowledgments}
    We thank S. Acharya for valuable discussions. This work was supported by the Air Force Office of Scientific Research under Award FA9550-22-1-0349 (D.A., I.Ž.), the National Science Foundation under Award DMR-2532768 (D.A., I.Ž.), by the National Natural Science Foundation of China Grant No. 12474155 (T.Z.), the Zhejiang Provincial Natural Science Foundation of China Grant No. LR25A040001 (T.Z.), and by the UB Center for Computational Research. K.B. acknowledges support from the National Science Foundation through Grant OIA-2521415 and from a UNL Grand Challenges catalyst award entitled Quantum Approaches Addressing Global Threats. The NRL portion of this work was authored by the National Laboratory of the Rockies for the U.S. Department of Energy (DOE) under Contract No. DE-AC36-08GO28308. Funding was provided by the Computational Chemical Sciences program within the Office of Basic Energy Sciences, U.S. Department of Energy. We acknowledge the use of the National Energy Research Scientific Computing Center, under Contract No. DE-AC02-05CH11231 using NERSC award BES-ERCAP0021783 and also that a portion of the research was performed using computational resources sponsored by the DOE's Office of Energy Efficiency and Renewable Energy and located at the  National Laboratory of the Rockies and computational resources provided by the Oak Ridge Leadership Computing Facility.  The views expressed in the article do not necessarily represent the views of the DOE or the U.S. Government.

\section*{Author Declarations}

\vspace{-0.2cm}

\subsection*{Conflict of Interest}

\vspace{-0.2cm}
    
The authors declare no conflict of interest.

\vspace{-0.6cm}
    
\subsection*{Author Contributions}
\textbf{Denzel Ayala}  Code Development; Conceptualization; Data Curation; Formal analysis; Investigation; Writing-Original Draft
\textbf{Dimitar Pashov}  Code Development; Data Curation; Investigation
\textbf{Tong Zhou} Conceptualization; Writing-review and editing
\textbf{Kirill Belashchenko} Conceptualization; Formal analysis; Writing-review and editing
\textbf{Mark van Schilfgaarde} Formal analysis; Funding acquisition;  Writing-review and editing; Supervision (Equal)
\textbf{Igor \v{Z}uti\'{c}} Conceptualization; Funding acquisition; Writing-review and editing; Supervision (Equal)

\section{Data Availability}
    The data that support the findings of this study are available within the article. The data were obtained from the open-source package Questaal, available at https://www.questaal.org.
\FloatBarrier
\appendix
\section{Convergence Information}

\begin{figure}[b]
\centering
    \includegraphics[width=\columnwidth]{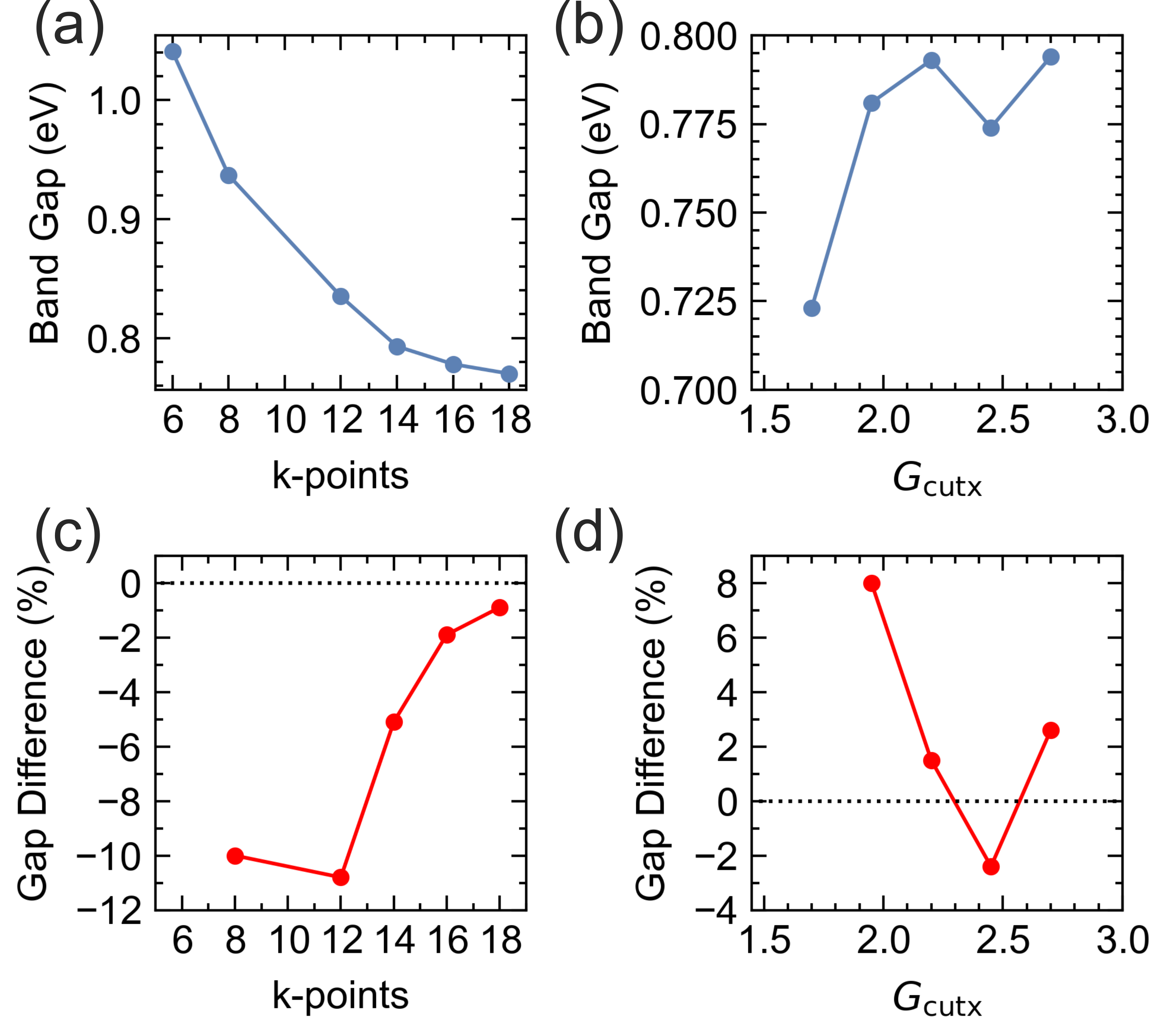}
    \caption{Testing QS$GW$ convergence of $E_g$ for FE 3R 2L $\alpha$-In$_2$Se$_3$. (a) Computed $E_g$ with a fixed $G_\text{cutx}=2.2$ and varying the k-points number in the plane. (b) Computed $E_g$ with fixed k-points, $14\times14\times1$. (c) The difference in $E_g$ as the k-points
    are varied, less than 2\% between 16 and 14 k-points, such that 
    14 k-points are used in presented figures.
    (d) The difference in $E_g$ as $G_\text{cutx}$ was varied.}
    \label{fig:SF1}
\end{figure}

    Within the \qsgw\ framework there are two types of self-consistency field (SCF) loops: (i) For the electron density and (ii) for quasiparticles. The first, DFT-level, iterates the one-particle equations to update $n(\mathbf{r})$\cite{Kohn1965:PR} and the corresponding $V_H$, for a fixed effective $V_\text{xc}$, while the latter, \qsgw-level, updates the nonlocal static effective potential $\Sigma_0$ until the quasiparticle energies and eigenfunctions are stable. The two SCF loops have different convergence parameters.

    For thin films, the \qsgw\ SCF loops require extensive convergence testing and the computational time increases with the vacuum thickness. For the k-point and G-vector cutoff convergence testing in Fig.~\ref{fig:SF1} the vacuum thickness was 40$\;$\AA. The parameters were varied until the difference of the computed $E_g$ was less than $\sim$5\%, see Figs.~\ref{fig:SF1}(c) and \ref{fig:SF1}(d). During the \qsgw\ k-point convergence testing, the DFT level k-points were set to be twice as many as the \qsgw\ k-points. The constant ratio minimized potential numerical artifacts. For the convergence testing of $G_\text{cutx}$, the k-points were fixed and  $G_\text{cutb} = G_\text{cutx} + 0.55$ Ry$^{1/2}$. The decrease in $E_g$ at $G_\text{cutx}=2.5$ Ry$^{1/2}$ is likely due to numerical instabilities as these $G_\text{cutx}$ values are very large for an $s-p$ system.

\begin{figure}[t]
\centering
    \includegraphics[width=\columnwidth]{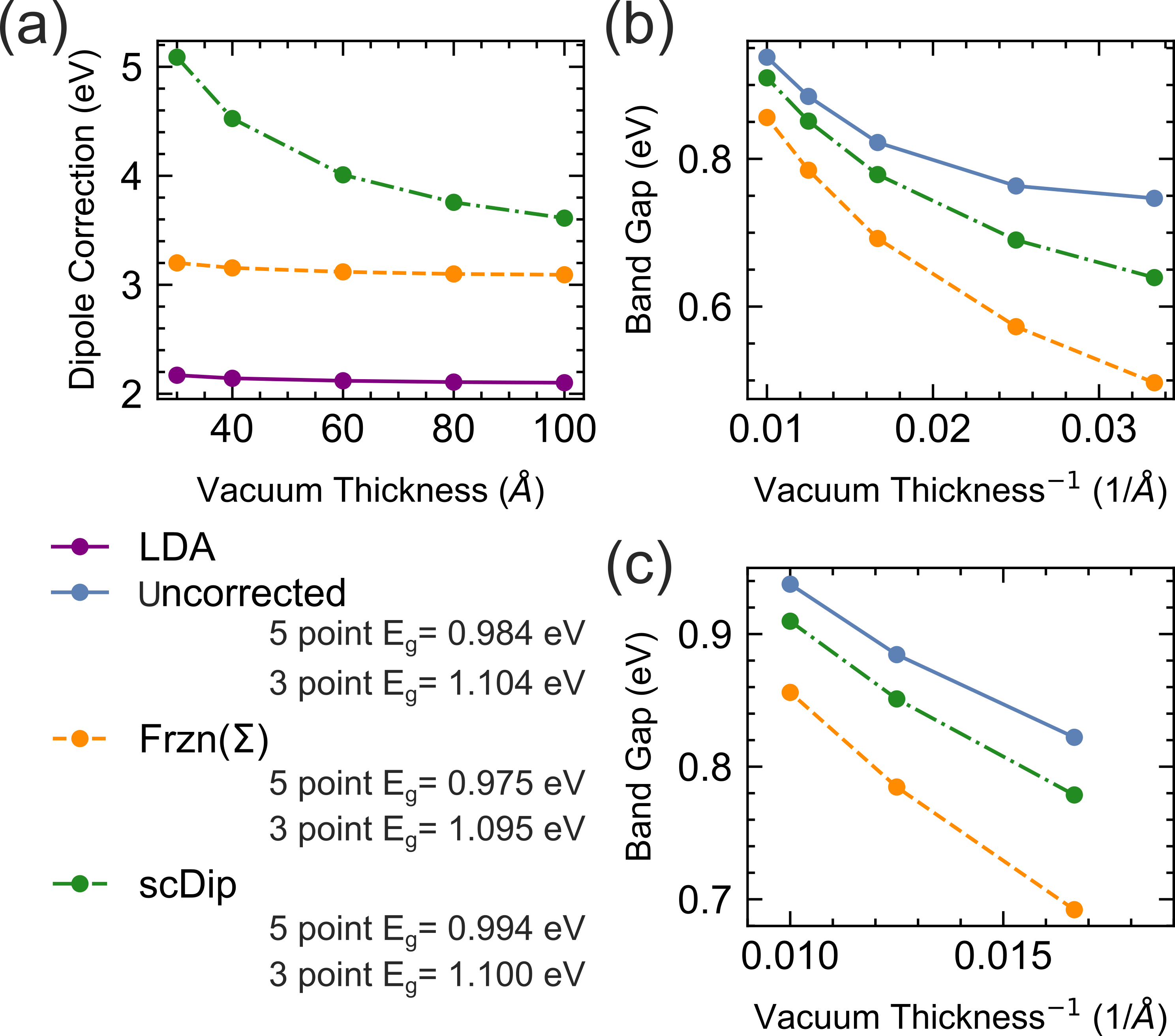}
    \caption{The vacuum thickness dependence of FE 3R 2L $\alpha$-\inse.  For the Frzn($\Sigma$) plots the self-energy, $\Sigma$, and electron density, $n(\mathbf{r})$, were well converged via QS$GW$ without a dipole correction. $n(\mathbf{r})$ was then updated to DFT-level self-consistency with dipole correction included while holding $\Sigma$ fixed. (a) The magnitude of the dipole correction for various approximations. (b)-(c) $E_g$ at various vacuum thicknesses. 
    }
    \label{fig:SF2}
\end{figure}   

    Within \qsgw, the DFT-level dipole correction of the Hartree potential can be included either as the frozen self-energy approach, Frzn($\Sigma$), or the self-consistent dipole inclusions approach, scDip. The dipole correction itself is a one-body term that was integrated into the many-body workflow to remove spurious contributions from the Hartree potential. There was no dipole correction capability prior to this work and these two methods are a practical first step in the full many-body implementation of the dipole correction. In the Frzn($\Sigma$) approach, Fig.~\ref{fig:SF2} (orange lines), the system has converged self-consistently at the \qsgw-level and the self-energy is then frozen while  $n(\mathbf{r})$ is re-converged with the dipole correction added in. This is a post processing step but has the weakness that when the self-energy is being constructed the fictitious inter-image interactions from $V_H$ are present. The scDip method, Fig.~\ref{fig:SF2} (green lines), simply includes the dipole correction during every step of the \qsgw\ cycle. While this approach corrects the one-particle inter-image interactions the calculations were very sensitive to the mixing parameters and convergence threshold. At smaller vacuum thicknesses the difference in computed properties between these two approaches was significant, Fig.~\ref{fig:SF2}. For the magnitude of the correction the difference at 30$\;$\AA\ of vacuum was $\sim2\;$eV while for $E_g$ it was $\sim0.5\;$eV.

    As the dipole correction only treats the one-particle component to the potential, $V_H$, the effect of the nonlocal $V_{xc}$ was determined by a vacuum thickness convergence. At smaller thicknesses, the change in the computed $E_g$ was nonlinear, Fig.~\ref{fig:SF2}(b). The linear fit for the gap was done with a reduced dataset, Fig.~\ref{fig:SF2}(c). By extrapolating to the infinite thickness, using a linear fit of the difference in $E_g$ between the uncorrected, Frzn($\Sigma$), and the full dipole correction became negligible. We  found that the magnitude of the dipole correction in Frzn($\Sigma$) method was less sensitive to the vacuum thickness.   

\bibliography{In2Se3}

\end{document}